\documentclass[aps,prx,twocolumn,nofootinbib,longbibliography]{revtex4-1}
\usepackage{bm, graphicx, amsmath, amssymb}
\usepackage{bbm}
\usepackage[mathscr]{eucal}
\usepackage{mathtools}
\usepackage{epsfig,graphics,graphicx}
\usepackage{xcolor}
\usepackage{lipsum}
\usepackage{enumitem}
\usepackage{ytableau}

\usepackage{hyperref}
\hypersetup{citecolor=blue}
\hypersetup{colorlinks=true}

\newcommand{\id}{\openone}
\newcommand{\tr}{{\rm tr}\,}
\newcommand{\ket}[1]{\left|{#1}\right\rangle}

\newcommand{\braket}[2]{\langle{#1}|{#2}\rangle}
\newcommand{\ketbrad}[1]{\left|{#1}\rangle\!\langle{#1}\right|}
\newcommand{\ketbra}[2]{\left|{#1}\rangle\!\langle{#2}\right|}

\newcommand{\norm}[1]{||#1||}

\newcommand{\blue}[1]{{#1}}

\usepackage{mathtools}

\DeclarePairedDelimiter\floor{\lfloor}{\rfloor}

\newcommand{\x}{{\bf x}}
\renewcommand{\r}{{\bf s}}

\newcommand{\la}{{(\lambda)}}
\newcommand{\lb}{{\{\lambda\}}}
\renewcommand{\l}{\lambda}
\newcommand{\sym}{{\rm sym}}

\begin{document}

\title{Unsupervised classification of quantum data}

\author{Gael Sent\'is,$^{1}$ Alex Monr\`as,$^{2}$  Ramon Mu\~noz-Tapia,$^{2}$ John Calsamiglia,$^{2}$ and Emilio Bagan$^{2,3}$}
\affiliation{$^{1}$Naturwissenschaftlich-Technische Fakult\"at, Universit\"at Siegen, 57068 Siegen, Germany\\
$^{2}$F\'{i}sica Te\`{o}rica: Informaci\'{o} i Fen\`{o}mens Qu\`antics, Departament de F\'{\i}sica, Universitat Aut\`{o}noma de Barcelona,08193 Bellaterra (Barcelona), Spain\\
$^{3}$Department of Computer Science, The University of Hong Kong, Pokfulam Road, Hong Kong
}

\begin{abstract}

We introduce the problem of unsupervised classification of quantum data, namely, of systems whose quantum states are unknown. We derive the optimal single-shot protocol 
for the  binary case, where the states in a disordered input array are of two types. Our protocol 
is universal and able to automatically sort the input under minimal assumptions, yet partially preserving information contained in the states.
We quantify analytically its performance for arbitrary size and dimension of the data. We contrast it with the performance of its classical counterpart, which clusters data that has been sampled from two unknown probability distributions. We find that the quantum protocol
fully exploits the dimensionality of the quantum data to achieve a much higher performance, provided data is at least three-dimensional. \blue{For the sake of comparison, we discuss the optimal protocol when the classical and quantum states are known.}

\end{abstract}

\maketitle

\section{Introduction}

Quantum-based communication and computation technologies promise unprecedented applications and unforeseen speed-ups for certain classes of computational problems. In origin, the advantages of quantum computing were exemplary showcased through instances of problems that are hard to solve in a classical computer, such as integer factorization~\cite{Shor1998}, unstructured search~\cite{Grover1997}, discrete optimization~\cite{Finnila1994,Kadowaki1998}, 
and simulation of many-body Hamiltonian dynamics~\cite{Lloyd1996}.
In recent times, the field has ventured one step further: quantum computers are now also envisioned as nodes in a network of quantum devices, where connections are established via quantum channels, and data are 
quantum systems that flow through the network~\cite{Kimble2008,Wehner2018}. The design of future quantum networks in turn brings up new theoretical challenges, such as devising universal information processing protocols optimized to work with generic quantum inputs, without the need of human intervention.

Quantum learning algorithms are by design well suited for this class of automated tasks~\cite{Dunjko2017}. Generalizing classical machine learning ideas to operate with quantum data, some algorithms have been devised for quantum template matching~\cite{Sasaki2002}, quantum anomaly detection~\cite{Liu2017,Skotiniotis2018}, learning unitary transformations~\cite{Bisio2010} and quantum measurements~\cite{Bisio2011a}, and classifying quantum states~\cite{Guta2010,Sentis2012a,Sentis2014a,Fanizza2018}. These works fall under the broad category of \emph{supervised} learning~\cite{Hastie2001,Devroye2013}, where the aim is to learn an unknown conditional probability distribution ${\rm Pr}(y|x)$ from a number of given samples $x_i$ and associated values or labels $y_i$, called \emph{training} instances. The performance of a trained learning algorithm is then evaluated by applying the learned function over new data $x'_i$ called \emph{test} instances. In the quantum extension of supervised learning~\cite{Monras2017}, the training instances are quantum---say, copies of the quantum state templates, or a potential anomalous state, or a number of uses of an unknown unitary transformation. The separation between training and testing steps is sometimes not as sharp: in reinforcement learning, training occurs on an instance basis via the interaction of an agent with an environment, and the learning process itself may alter the underlying probability distribution~\cite{Dunjko2016}.

In contrast, \emph{unsupervised} learning aims at inferring structure in an unknown distribution ${\rm Pr}(x)$ given random, unlabeled samples $x_i$. Typically, this is done by grouping the samples in \emph{clusters}, according to a preset definition of similarity. Unsupervised learning is a versatile form of learning, attractive in scenarios where 
appropriately labeled training data is not available or too costly.
But it is also---generically---a much more challenging problem \cite{Aloise2009,Ben-David2015}. To our knowledge, a quantum extension of unsupervised learning in the sense described above
has not yet been considered in the literature. 
\begin{figure}[t]
	\includegraphics[scale=.6]{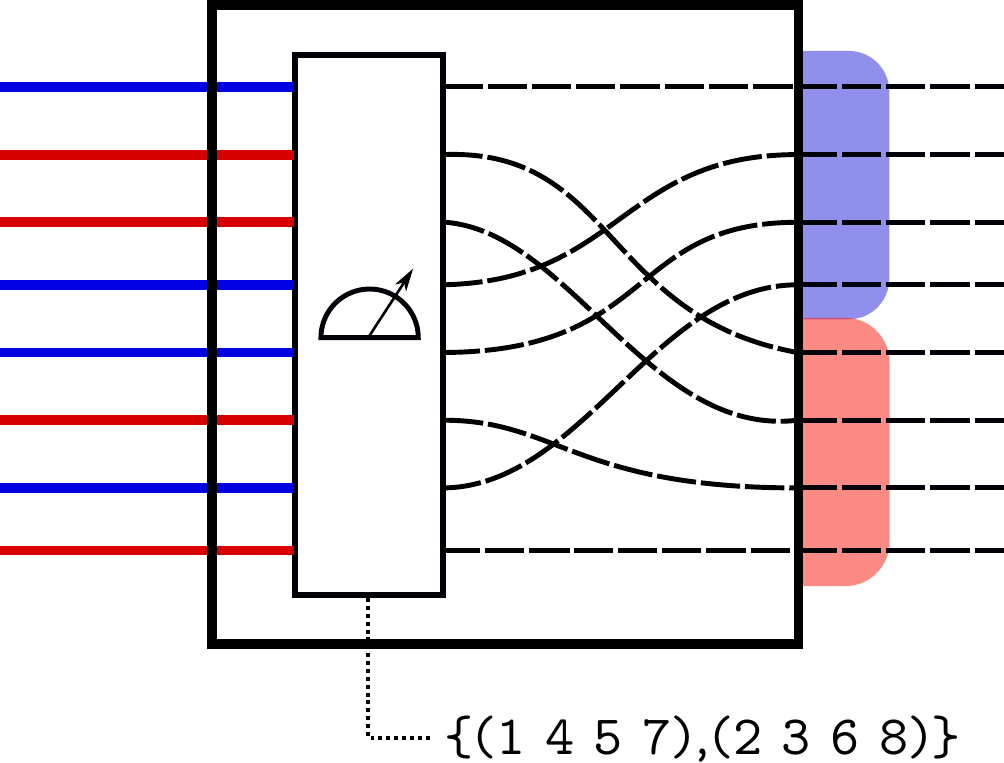}
	\caption{Pictorial representation of the clustering device for an input of eight quantum states. States of the same type have the same color. States are clustered according to their type by performing a suitable collective measurement, which also provides a classical description of the clustering.}\label{fig:scheme}
\end{figure}
In this paper, we take a first step into this branch of quantum learning by introducing the problem of unsupervised binary classification of quantum states. We consider the following scenario: a source prepares quantum systems in two possible pure states that are completely unknown; after some time, $N$ such systems have been produced and we ask ourselves whether there exists a quantum device that is able to cluster them in two groups according to their states (see Fig.~\ref{fig:scheme}). 
\blue{This scenario represents a quantum clustering task in its simplest form,
where the single feature defining a cluster of quantum systems is that their states are identical. 
While clustering classical data under this definition of cluster---a set of equal data instances---yields a trivial algorithm, 
merely observing such simple feature in a finite quantum data set 
involves a nontrivial stochastic process
and gives rise to a primitive of operational relevance for quantum information. 
Moreover, in some sense our scenario actually contains a
classical binary clustering problem: if we were to measure each quantum system separately, we would obtain a set of $N$ data points (the measurement outcomes). The points would be effectively sampled from the two probability distributions determined by the quantum states and the choice of measurement. The task would then be to identify which points were sampled from the same distribution.
Reciprocally, we can interpret our quantum clustering task as a natural extension of a classical clustering problem with completely unstructured data, where the only single feature that identifies a cluster is that the data points are sampled from a fixed, but arbitrary, categorical probability distribution (i.e., with no order nor metric in the underlying space). The quantum generalization is then to consider (non-commuting) quantum states instead of probability distributions.}

We require two important features in our quantum clustering device: (i) it has to be universal, that is, it should be designed to take any possible pair of types of input states, and (ii) 
it has to provide a classical description of the clustering, that is, which particles belong to each cluster.
Feature (i) ensures 
general purpose use and versatility of the clustering device, in a similar spirit to programmable quantum processors~\cite{Buzek2006}. Feature (ii) allows us to assess the performance of the device purely in terms of the accuracy of the clustering, which in turn facilitates the comparison with classical clustering strategies. Also due to (ii), we can justifiably say that the device has not only performed the clustering task but also ``learned'' that the input is (most likely) partitioned as specified by the output description. 
Note that relaxing 
feature (ii) in principle opens the door to a more general class of \emph{sorting} quantum devices, where the goal could be, e.g., to minimize the distance (under some norm) between the global output state and the state corresponding to perfect clustering of the input. Such devices, however, fall beyond the scope of unsupervised learning.

Requiring the description of the clusters as a classical outcome induces structure in the device. To generate this information, a quantum measurement shall be performed over all $N$ systems with as many outcomes as possible clusterings. Then, the systems will be sorted according to this outcome (see Fig.~\ref{fig:scheme}). 
Depending on the context, e.g., on whether or not the systems will be further used after the clustering, different figures of merit shall be considered in the optimization of the device. 
In this paper we focus on the clustering part: our goal is to find the quantum measurement that maximizes the success probability of a correct clustering. 

\blue{
Features (i) and (ii) allow us to formally regard quantum clustering as a state discrimination task~\cite{Helstrom1976,Barnett2001,Chiribella2004,Chiribella2006a,Audenaert2007,Krovi2015a}, 
albeit with important differences with respect to the standard setting. In quantum state discrimination~\cite{Helstrom1976},
we want to determine the state of a quantum system among a set of \emph{known} hypotheses (i.e., classical descriptions of quantum states).
We can phrase this problem in machine learning terminology as follows. We have a test state (or several copies of it~\cite{Audenaert2007}) and we decide its label based on \emph{infinite training} data. In other words, we have full knowledge about the meaning of the possible labels. Supervised quantum learning algorithms for quantum state classification~\cite{Guta2010,Sentis2012a,Sentis2014a,Fanizza2018} consider the intermediate scenario with \emph{limited training} data. In this case,  no description of the states is available. Instead, we are provided with a finite number of copies of systems in each of the possible quantum states, and thus we have only partial classical knowledge about the labels. Extracting the label information from the quantum training data then becomes a key step in the protocol. Following this line of thought, the problem we consider in this paper is a type of unsupervised learning, that is, one with \emph{no training}. There is no information whatsoever about what state each label represents.
}

We obtain analytical expressions for the performance of the optimal clustering protocol for arbitrary values of the local dimension $d$ of the systems in the cases of finite number of systems $N$ and in the asymptotic limit of many systems. We show that, in spite of the fact that the number of possible clusterings grows exponentially with $N$, the success probability decays only as $O(1/N^2)$. 
Furthermore, we contrast these results with an optimal clustering algorithm designed for the classical version of the task. 
We observe a striking phenomenon when analyzing the performance of the two protocols for $d>2$: whereas increasing the local dimension has a rapid negative impact in the success probability of the classical protocol (clustering becomes, naturally, harder), it turns out to be beneficial for its quantum counterpart. 

We also see, through numerical analysis, that the quantum measurement that maximizes the success probability is also optimal for a more general class of cost functions that are more natural for clustering problems, including the Hamming distance. 
In other words, this provides evidence that our entire analysis does not depend strongly on the chosen figure of merit, but rather on the structure of the problem itself.

Measuring the systems will in principle degrade the information encoded in their states, hence, intuitively, there should be a trade-off between how good a clustering is and how much information about the original states is left in the clusters. Remarkably, our analysis reveals that the measurement that clusterizes optimally actually preserves information regarding the type of states that form each cluster.  
\blue{This feature adds to the usability of our device as a universal quantum data sorting processor. It can be regarded as the quantum analogue of a sorting network (or sorting memory)~\cite{Knuth1998}, used as a fixed network architecture that automatically orders generic inputs coming from an aggregated data pipeline.}	
The details of this second step are however left for a subsequent publication.

The paper is organized as follows. 
In Section~\ref{sec:the_task}, we formalize the problem and derive the optimal clustering protocol and its performance. In Section~\ref{sec:classical}, we consider a classical clustering protocol and contrast it with the optimal one. \blue{We present the proofs of the main results of our work and the necessary theoretical tools to derive them in Section~\ref{sec:methods}.}
We end in Section~\ref{sec:discussion} discussing the features of our quantum clustering device and other cost functions, and giving an outlook on future extensions.

\section{Clustering quantum states}\label{sec:the_task}

Let us suppose that a source prepares quantum systems randomly in one of two pure $d$-dimensional states~$\ket{\phi_0}$ and~$\ket{\phi_1}$ with equal prior probabilities. 
Given a sequence of $N$ systems produced by the source, and with no knowledge of the states~$\ket{\phi_{0/1}}$, we are required to assign labels~`0' or~`1' to each of the systems. The labeling can be achieved via a generalized quantum measurement that tries to distinguish among all the possible global states of the $N$ systems. Each outcome of the measurement will then be associated to a possible label assignment, that is, to a \emph{clustering}.

\begin{figure}[t]
	\includegraphics[scale=.35]{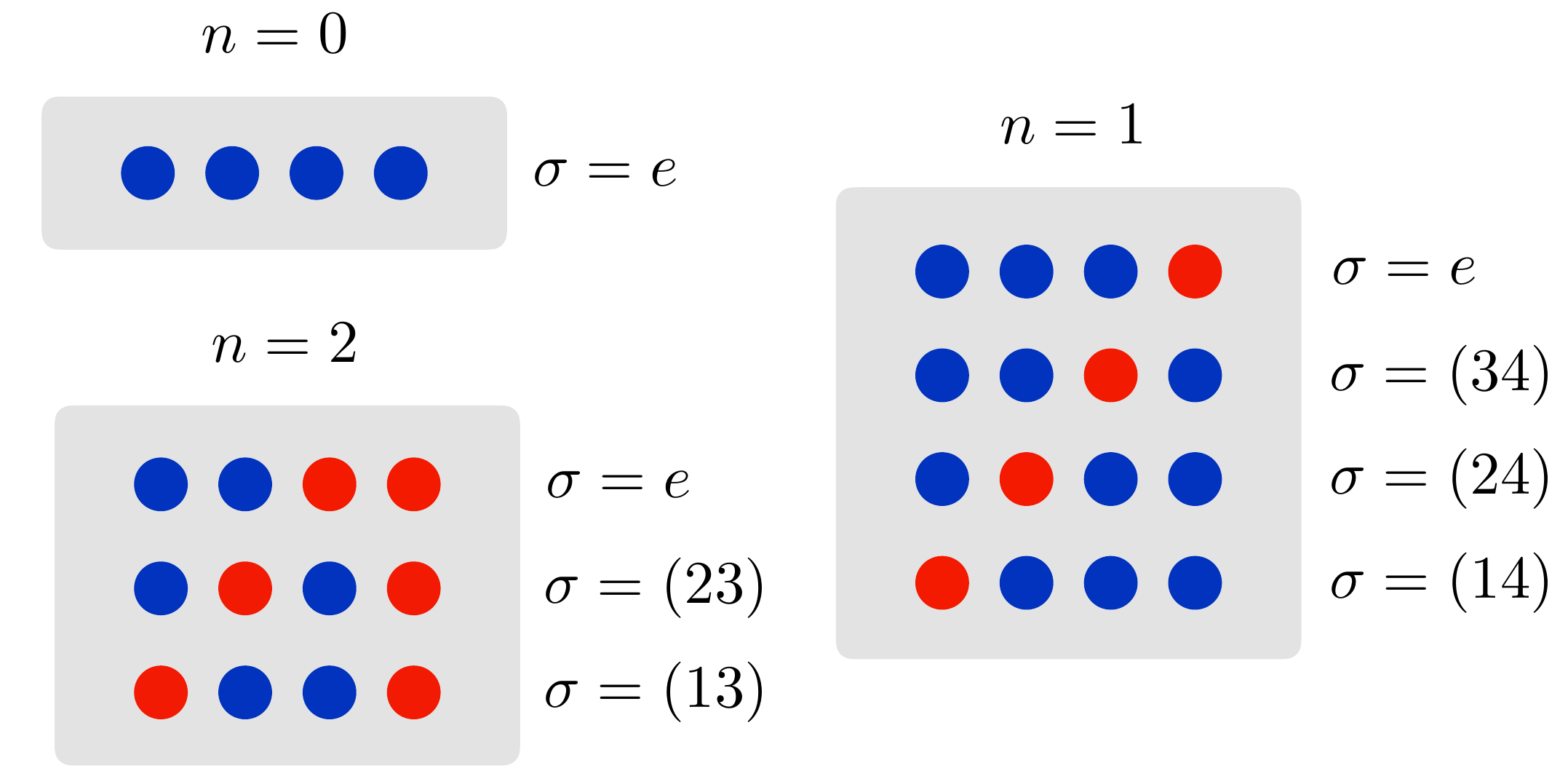}
	\caption{All possible clusterings of $N=4$ systems when each can be in one of two possible states, depicted as blue and red. The pair of indices $(n,\sigma)$ identifies each clustering, where $n$ is the size of the smallest cluster, and~$\sigma$ is a permutation of the {\em reference} clusterings (those on top of each box), wherein the smallest cluster falls on the right. The symbol $e$ denotes the identity permutation, and $(ij)$ the transposition of systems in positions $i$ and $j$. Note that the choice of $\sigma$ is not unique.}\label{fig:N4}
\end{figure}

Consider the case of four systems. All possible clusterings that we may arrange are depicted in Fig.~\ref{fig:N4} as strings of red and blue balls. Since the individual states of the systems are unknown, what is labeled as ``red'' or ``blue'' is arbitrary, thus interchanging the labels leads to an equivalent clustering. For arbitrary $N$, there will be~$2^{N-1}$ such clusterings. Fig.~\ref{fig:N4} also illustrates a natural way to label each clustering as $(n,\sigma)$. The index~$n$ counts the number of systems in the smallest cluster. The index $\sigma$ is a permutation that brings a \emph{reference} clustering, defined as that in which the systems belonging to the smallest cluster fall all on the right, into the desired form. To make this labeling unambiguous, $\sigma$ is chosen from a restricted set ${\mathscr S}_n\subset S_N$, where $S_N$ stands for the permutation group of $N$ elements and $e$ denotes its unity element. We will see that the optimal clustering procedure consists in measuring first the value of $n$, and, depending on the outcome, performing a second measurement that identifies $\sigma$ among the relevant permutations with a fixed $n$. 

Thus, unsupervised clustering has been cast as a  multi-hypothesis discrimination problem, which can be solved for an arbitrary number of systems $N$ with local dimension $d$. Below, we outline the derivation of our main result: the expression of the maximum average success probability achievable by a quantum clustering protocol. In the limit of large $N$ \blue{and for arbitrary $d$ (not necessarily constant with $N$)}, 
we show that this probability behaves as\footnote{\blue{The symbol $\sim$ stands for ``asymptotically equivalent to'', as in~\citep{Abramowitz1965}.}}
\begin{equation}\label{ps_asym}
P_{\rm s} \sim {8(d-1)\over\left(\displaystyle 2d+N\right) N}\,.
\end{equation}
Naturally, $P_{\rm s}$ goes to zero with $N$, since the total number of clusterings increases exponentially and it becomes much harder to discriminate among them. What may perhaps come as a surprise is that, despite this exponential growth, the scaling of $P_{\rm s}$ is only of order $O(1/N^2)$.\footnote{\blue{It is also interesting to see how far can one improve this result. By letting $d$ scale with $N$, e.g., by substituting $d\sim s N^\gamma$ for some $s>0$, $\gamma>1$ in Eq.~\eqref{ps_asym}, we obtain the absolute maximum
$P_{\rm s} \sim 4/N$.}} 
Furthermore, increasing the local dimension yields a linear improvement in the asymptotic success probability. As we will later see, whereas the asympotic behavior in $N$ is not an exclusive feature of the optimal quantum protocol---we observe the same scaling in its classical counterpart, albeit only when $d=2$---the ability to exploit extra dimensions to enhance distinguishability is.

Let us present an outlined derivation of the optimal quantum clustering protocol. Each input can be described by a string of 0's and 1's ${\x}=(x_1\cdots x_N)$, so that the global state of the systems entering the device is
$\ket{\Phi_\x} = \ket{\phi_{x_1}}\otimes\ket{\phi_{x_2}}\otimes\cdots\otimes \ket{\phi_{x_N}}$.
The clustering device can generically be defined by a positive operator valued measure (POVM) with elements $\{E_\x\}$, fulfilling $E_\x\ge 0$ and $\sum_\x E_\x = \openone$, where each operator~$E_\x$ is associated to the statement ``the measured global state corresponds to the string $\x$''. We want to find a POVM that maximizes the average success probability $P_{\rm s}=2^{1-N}\int d\phi_0 d\phi_1 \sum_\x \tr (\ketbrad{\Phi_\x} E_\x)$, where we assumed that each clustering is equally likely at the input, and we are averaging over all possible pairs of states $\{\ket{\phi_{0}},\ket{\phi_{1}}\}$ and strings $\x$. Since our goal is to design a universal clustering protocol, the operators $E_\x$ cannot depend on $\ket{\phi_{0,1}}$, and we can take the integral inside the trace. The clustering problem can then be regarded as the optimization of a POVM that distinguishes between effective density operators of the form
\begin{equation}\label{rhox}
\rho_{\x}= \int d\phi_0 \, d\phi_1 \ketbrad{\Phi_\x} \,.
\end{equation}
It now becomes apparent that $\rho_\x=\rho_{\bar \x}$, where $\bar \x$ is the complementary string of $\x$ (i.e., the values 0 and 1 are exchanged).

The key that reveals the structure of the problem and allows us to deduce the optimal clustering protocol resides in computing the integral in Eq.~\eqref{rhox}. Averaging over the states leaves out only the information relevant to identify a clustering, that is, $n$ and $\sigma$. Certainly, identifying $\x\equiv(n,\sigma)$, we can rewrite $\rho_\x$ as
\begin{align}\label{rho_ns}
\rho_{n,\sigma} &= c_n \, U_\sigma \, (\openone^\sym_n\otimes\openone^\sym_{N-n} )\,U_\sigma^\dagger \nonumber\\
 &= c_n \bigoplus_\l \id_\la \otimes \Omega_\lb^{n,\sigma}\,.
\end{align}
By applying Schur lemma, one readily obtains the first line, where $\openone^\sym_k$ is a projector onto the completely symmetric subspace of $k$ systems, 
$c_n$ is a normalization factor, 
and $U_\sigma$ is a unitary matrix representation of $\sigma$. The second line follows from using the Schur basis (see Section~\ref{app:optimality}), in which the states  $\rho_{n,\sigma}$ are block-diagonal.
Here $\l$ labels the irreducible representations---irreps for short---of the joint action of the groups ${\rm SU}(d)$ and $S_N$ over the vector space $(d,\mathbb C)^{\otimes N}$, and is usually identified with the shape of Young diagrams (or partitions of~$N$).  
A pair of parentheses, $()$ [brackets, $\{\}$], surrounding the subscript $\lambda$, e.g., in Eq.~\eqref{rho_ns},  are used when 
$\lambda$ refers exclusively to irreps of ${\rm SU}(d)$ [$S_N$];   
we stick to this convention throughout the paper.
Note that averaging over all ${\rm SU}(d)$ transformations erases the information contained in the representation subspace~$(\lambda)$. 
It also follows from Eq.~\eqref{rho_ns} and the rules of the Clebsch-Gordan decomposition that (i)~only two-row Young diagrams (partitions of length two) show up in the direct sum above, and (ii)~the operators $\Omega_\lb^{n,\sigma}$ are rank-1 projectors (see Appendix~\ref{app:irreps}). They carry all the information relevant for the clustering, and are understood to be zero for irreps $\l$ outside the support of~$\rho_{n,\sigma}$.

With Eq.~\eqref{rho_ns} at hand, the optimal clustering protocol can be succinctly described as two successive measurements---we state the result here and present an optimality proof in Section~\ref{app:optimality}. The first measurement is a projection onto the irrep subspaces $\l$,
described by the set $\{\openone_\la \otimes \openone_\lb\}$.
The outcome of this measurement provides an estimate of~$n$, as $\l$ is one-to-one related to the size of the clusters. More precisely, we have from~(i) that $\l=(\l_1,\l_2)$, where $\l_1$ and $\l_2$ are nonnegative integers such that $\l_1+\l_2=N$ and~$\l_1\ge \l_2$. 
Then, given the outcome $\l=(\l_1,\l_2)$ of this first measurement, the optimal guess turns out to be $n=\l_2$.
Very roughly speaking, the ``asymmetry" in the subspace $\l=(\l_1,\l_2)$ increases with $\l_2$. 
We recall that $\l=(N,0)$ is the fully symmetric subspace of $(d,\mathbb{C})^N$. Naturally, $\rho_{0,\sigma}$ has support only in this subspace, as all states in the data are of one type. 
As $\l_2$ increases from zero, more states of the alternative type are necessary to achieve the increasing asymmetry of $\l=(\l_1,\l_2)$. 
Hence, for a given $\l_2$, there is a minimum value of $n$ for which~$\rho_{n,\sigma}$ can have support in the subspace $\l=(\l_1,\l_2)$. This minimum $n$ is the optimal guess.

Once we have obtained a particular $\l\!=\!\l^*$ as an outcome (and guessed $n$), a second measurement is performed over the subspace $\{\l^*\}$ to produce a guess for~$\sigma$. 
Since the states $\rho_{n,\sigma}$ are covariant under~$S_N$, the optimal measurement to guess the permutation~$\sigma$
is also covariant,
and its seed is the rank-1 operator $\Omega_{\{\l^*\}}^{n, e}$, where $\l^*= (N-n,n)$. Put together, these two successive measurements 
yield a joint optimal POVM whose elements take the form 
\begin{equation}\label{povm_elements_main}
E_{n,\sigma} = \xi_{\l^*}^{n} (\openone_{(\lambda^*)} \otimes \Omega_{\{\lambda^*\}}^{n,\sigma})\,,
\end{equation}
where $(n,\sigma)$ is the guess for the cluster 
and $\xi_{\l^*}^n$  
is some coefficient that guarantees  the POVM condition \mbox{$\sum_{n,\sigma} E_{n,\sigma}=\openone$}. 

The success probability of the optimal protocol can be computed as~$P_{\rm s}=2^{1-N}\sum_{n,\sigma} \tr(\rho_{n,\sigma} E_{n,\sigma})$ (see Section~\ref{app:optimality}). It reads
\begin{align}
P_{\rm s} &= 2^{1-N} \sum_{i=0}^{\floor{N/2}} \binom{N}{i} \frac{(d-1)(N-1-2i)^2}{(N-1+d-i)(i+1)^2} \,, \label{ps}
\end{align}
from which the asymptotic limit Eq.~\eqref{ps_asym} follows (see Appendix~\ref{app:asymptotics}).

\blue{ Before closing this section we would like to briefly discuss the case when some information about the possible states $\ket{\phi_0}$ and $\ket{\phi_1}$ is available. A clustering device that incorporates this information into its design should succeed with a probability higher than Eq.~\eqref{ps}, at the cost of universality. To explore the extent of this performance enhancement, we study the extreme case where we have full knowledge of the states $\ket{\phi_0}$ and $\ket{\phi_1}$. We find that in the large $N$ limit the maximum improvement is by a factor of $N$. The optimal success probability scales as
\begin{equation}\label{ps_known}
P_{\rm s} \sim \frac{4(d-1)}{N} 
\end{equation}
(see Section~\ref{sec:quantumknown} for details).
}

\section{Clustering classical states}\label{sec:classical}

To grasp the significance of our quantum clustering protocol, a comparison with a classical analogue is called for. 
First, in the place of a quantum system whose state is either $\ket{\phi_0}$ or $\ket{\phi_1}$, 
an input would be an instance of a $d$-dimensional random variable sampled from either one of two categorical 
probability distributions, $P=\{p_s\}_{s=1}^d$ and $Q=\{q_s\}_{s=1}^d$.
Then, given a string of samples $\r=(s_1\cdots s_N)$, $s_i\in\{1,\ldots,d\}$, the clustering task would consist in grouping the data points $s_i$ in two clusters so that all points in a cluster have a common underlying probability distribution. 

Second, in analogy with the quantum protocol, our goal would be to find the optimal universal  (i.e., independent of $P$ and $Q$) protocol, that performs this task. Here, optimality means attaining 
the maximum average success probability, where the average is over all $N$-length sequences $\x$ of distributions $P$ and $Q$ from which the string $\r$ is sampled,  and over all such distributions.

It should be emphasized that this is a very hard classical clustering problem, with absolute minimal assumptions, where there is no metric in the domain of the random variables and, in consequence, no exploitable notion of distance. Therefore, \blue{one should expect} the optimal algorithm to have a rather low performance and to differ significantly from well-known algorithms for classical unsupervised classification problems.

As a further remark, we note that a choice of prior is required to perform  the average over $P$ and $Q$. We will assume that the two are uniformly distributed over the simplex on which they are both defined. This reflects our lack of knowledge about the distributions underlying the string of samples $\r$. 

\blue{Under all these specifications, the classical clustering problem we just defined naturally connects with the quantum scenario in Section~\ref{sec:the_task} as follows. 
We can interpret 
$\r$ 
as 
a string of outcomes obtained upon performing the same projective measurement on each individual quantum state $|\phi_{x_i}\rangle$ of our original problem.
Furthermore, such local measurements can also be interpreted as a decoherence process affecting the pure quantum states at the input, whereby they decay into classical probability distributions over a fixed basis.} 
We might think of this as the semiclassical analogue of our original problem, since quantum resources are not fully exploited.

Let us first lay out the problem in the special case of $d=2$, where the underlying distributions are Bernoulli, and we can write  $P=\{p,1-p\}$, $Q=\{q,1-q\}$. Given an $N$-length string of samples~$\r$, our intuition tells us that the best we can do is to assign the same underlying probability distribution to equal values in~$\r$. So if, e.g., $\r=(00101\cdots)$, we will guess that the underlying sequence of distributions is 
$\hat\x=(PPQPQ\cdots)$ [or, equivalently, the complementary sequence $\hat\x=(QQPQP\cdots)$]. Thus, data points will be clustered according to their value 0 or 1. The optimality of this guessing rule is a particular case of the result for $d$-dimensional random variables in Appendix~\ref{app:classic}.

The probability that a string of samples $\r$, with $l$ zeros and $N-l$ ones, arises from the guessed sequence $\hat\x$ is given by
\begin{equation}
\!{\rm Pr}(\r|\x\!=\!\hat\x) \!=\!\!\!\int_0^1\!\! \!dp\!\!\int_0^1\!\!\! dq\,p^l q^{N-l} =\frac{1}{(l+1)(N-l+1)} \,.
\end{equation}
The average success probability can then be readily computed as 
$P_{\rm s}^{\rm cl}=2 \sum_{\x,\r} \delta_{\x,\hat\x}\,{\rm Pr}(\x)\,{\rm Pr}(\r|\x)$ (recall that $\hat \x$ depends on $\r$), where ${\rm Pr}(\x)=2^{-N}$ is the prior probability of the sequence $\x$, which we assume to be uniform. The factor $2$ takes into account that guessing the complementary sequence leads to the same clustering.  It is now quite straightforward to derive the asymptotic expression of $P_{\rm s}^{\rm cl}$ for large $N$. In this limit~$\x$ will typically have the same number of $P$ and $Q$ distributions, so the guess $\hat\x$ will be right if $l=N/2$. 
Then,
\begin{equation}
P_{\rm s}^{\rm cl} \sim 2 \frac{1}{(N/2+1)^2} \sim \frac{8}{N^2} \,.
\end{equation}
This expression coincides with the quantum asymptotic result in Eq.~\eqref{ps_asym} for $d=2$. As we now see, this is however a particularity of Bernoulli 
distributions.

The derivation for $d>2$ is more involved, since the optimal guessing rule is not so obvious (see Appendix~\ref{app:classic} for details). Loosely speaking, we should still assign samples with the same value to the same cluster. By doing so, we obtain up to $d$ preliminary clusters. We next merge them into two clusters  in such a way that their final sizes are as balanced as possible. This last step, known as the {\em partition problem}~\cite{Korf1998}, is weakly NP-complete. Namely,
\blue{its complexity is polynomial in the magnitudes of the data involved (the size of the preliminary clusters, which depends on $N$) but non-polynomial in the input size (the number of such clusters, determined by $d$).}
This means that the classical and semiclassical protocols cannot be implemented efficiently for arbitrary~$d$. In the asymptotic limit of large $N$, and for arbitrary fixed values of $d$, we obtain
\begin{equation}\label{ps_asym_cl}
P_{\rm s}^{\rm cl} \sim \left(\frac{2}{N}\right)^d \frac{(2d-2)!}{(d-2)!} \,.
\end{equation}
There is a huge difference between this result and Eq.~\eqref{ps_asym}. Whereas increasing the local dimension provides an asymptotic linear advantage in the optimal quantum clustering protocol---states become more orthogonal---it has the opposite effect in its classical and semiclassical analogues, as it reduces exponentially the success probability. 

In the opposite regime, i.e., for $d$ asymptotically large and fixed values of~$N$, the optimal classical and semiclassical strategies provide no improvement over random guessing, and the clustering tasks become exceedingly hard and somewhat uninteresting. This follows from observing that the guessing rule relies on grouping repeated data values. In this regime, the typical string of samples~$\r$ has no repeated elements, thus we are left with no alternative but to randomly guess the right clustering of the data and $P_{\rm s}^{\rm cl} \sim 2^{1-N}$.

\blue{
To complete the picture, we end up this section by considering known classical probability distributions. Akin to the quantum case, one would expect an increase in the success probability of clustering. An immediate consequence of knowing the distributions $P$ and $Q$ is that the rule for assigning a clustering given a string of samples~$\r$ becomes trivial. Each symbol $s_i\in\{1,\ldots,d\}$ will be assigned to the most likely distribution, that is, to $P$ ($Q$) if $p_{s_i} > q_{s_i}$ ($p_{s_i} < q_{s_i}$). It is clear that knowing $P$ and~$Q$ helps to better classify the data. 
This becomes apparent by considering the example of two three-dimensional distributions and the data string  $\r=(112)$. If the distributions are unknown, such sequence leads to the guess $\hat\x=(PPQ)$ [or equivalently to $\hat\x=(QQP)$]. In contrast, if $P$ and $Q$ are known and, e.g., $p_1>q_1$ and $p_2>q_2$, the same sequence leads to the better guess $\hat\x=(PPP)$. The advantage of knowing the distribution, however, vanishes in the large $N$ limit, and 
the asymptotic performance of the optimal clustering algorithm is shown to be given by Eq.~\eqref{ps_asym_cl}. The interested reader can find the details of the proof in  Appendix~\ref{app:classic_known}.
}

\blue{
\section{Methods}\label{sec:methods}

Here we give the full proof of optimality of our quantum clustering protocol/device, which leads to our main result in Eq.~\eqref{ps_asym}. The proof relies on representation theory of the special unitary and the symmetric groups. In particular, the Schur-Weyl duality is used to efficiently represent the structure of the input quantum data and the action of the device. We then leverage this structure to find the optimal POVM and compute the minimum cost. Basic notions of representation theory that we use in the proof are covered in the Appendices~\ref{app:partitions} and \ref{app:irreps}.
We close the Methods section proving Eq.~\eqref{ps_known} for the optimal success probability of clustering known quantum states.

\subsection{Clustering quantum states: unknown input states}\label{app:optimality}

In this
Section
we obtain the optimal POVM for quantum clustering and compute the minimum cost. First, we present a formal optimality proof for an arbitrary cost function $f(\x,\x')$, which specifies the penalty for guessing $\x$ if the input is $\x'$. Second, we particularize to the case of success probability, as discussed in the main text, for which explicit expressions are obtained.

\subsubsection{Generic cost functions} 

We say a POVM is optimal if it minimizes the average cost
\begin{equation}\label{app:av_cost}
{\bar f} = \int d\phi_0 \,d\phi_1 \sum_{\x,\hat\x} \eta_{\x} 
\, f(\x,\hat\x) \, {\rm Pr}(\hat\x|\x)
\,,
\end{equation}
where $\eta_{\x}$ is the prior probability of input string $\x$, and ${\rm Pr}(\hat\x|\x)=\tr (\ketbrad{\Phi_{\x}} E_{\hat\x}) $ is the probability of obtaining measurement outcome (and guess) $\hat\x$ given input $\x$; recall that $\ket{\Phi_\x} = \ket{\phi_{x_1}}\otimes\ket{\phi_{x_2}}\otimes\cdots\otimes \ket{\phi_{x_N}}$, $x_k=0,1$, and an average is taken over all possible pairs of states $\{\ket{\phi_{0}},\ket{\phi_{1}}\}$, hence $\x$ and its complementary $\bar\x$ define de same clustering. 
A convenient way to identify the different clusterings is by counting the number $n$, $0\leq n\leq \floor{N/2}$, of zeros in~$\x$ (so, strings with more 0s than 1s are discarded) and  giving a unique representative $\sigma$ of the equivalence class of permutations that turn the reference string $(0^n1^{\bar n})$, ${\bar n}=N-n$, into $\x$. We will denote the subset of these representatives by ${\mathscr S}_n\subset S_N$, and the number of elements in each equivalence class~by~$b_n$. A simple calculation gives us $b_n=2(n!)^2$ if $n=\bar n$, and $b_n=n!\bar n!$ otherwise. 

As discussed in the  main text, the clustering problem above is equivalent to a multi-hypothesis discrimination problem, where the hypotheses are given by
\begin{align}\label{app:rho_ns}
\rho_{\x} &= \int d\phi_0 \, d\phi_1 \ketbrad{\Phi_\x} \nonumber\\
&= c_n \, U_\sigma \, (\openone^\sym_n\otimes\openone^\sym_{\bar n} )\,U_\sigma^\dagger \,,
\end{align}
and we have used Schur lemma to compute the integral. Here,~$U_\sigma$ is a unitary matrix representation of the permutation~$\sigma$, $\openone^\sym_k$ is a projector onto the completely symmetric subspace of $k$ systems, and $c_n= 1/(D^\sym_n D^\sym_{\bar n})$, where~$D^\sym_k=s_{(k,0)}$ [see Eq.~(\ref{mult_s})] is the dimension of symmetric subspace of~$k$ qudits.

The states~\eqref{app:rho_ns} are block-diagonal in the Schur basis, which decouples the commuting actions of the groups ${\rm SU}(d)$ and $S_N$ over product states of the form of~$|\Phi_{\x}\rangle$. More precisely, Schur-Weyl duality states that the representations of the two groups acting on the common space $(d,\mathbb{C})^{\otimes N}$ are each other's commutant. Moreover,  it provides a decomposition of 
this space
into decoupled subspaces associated to irreducible representations (irreps) of both ${\rm SU}(d)$ and $S_N$. We can then express  the states~$\rho_{\x}$, where $\x$ is specified as $(n,\sigma)$ [$\x=(n,\sigma)$ for short], in the Schur basis as
\begin{equation}\label{app:rho_block}
\rho_{n,\sigma} = c_n \bigoplus_\l \id_\la \otimes \Omega_\lb^{n,\sigma} \,.
\end{equation}
In this direct sum, $\l$ is a label attached to the irreps of the joint action of ${\rm SU}(d)$ and $S_N$ and is usually identified with a partition of~$N$ or, equivalently, a Young diagram. As explained in the main text, a pair of parenthesis surrounding this type of label, like in $\la$, mean that it refers specifically to irreps of ${\rm SU}(d)$. Likewise, a pair of brackets, e.g., $\lb$, indicate that the label refers to irreps of~$S_N$. In accordance with this convention,  Schur-Weyl duality implies that $\Omega_\lb^{n,\sigma}=U_\sigma^\l \,\Omega_\lb^{n,e}\, (U_\sigma^\l)^\dagger$, where~$U_\sigma^\l$ is the matrix of the irrep~$\l$ that represents $\sigma\in S_N$, and $e$ denotes the identity permutation (for simplicity, we omit the index $e$ when no confusion arises). 
In other words, the family of states~$\rho_{n,\sigma}$ is covariant with respect to $S_N$. One can easily check that $\Omega_\lb^{n,\sigma}$ is always a rank-1 projector (see Appendix~\ref{app:irreps}).  In Eq.~\eqref{app:rho_block} it is understood that $\Omega_\lb^{n,\sigma}=0$ outside of the range of~$\rho_{n,\sigma}$.

With no loss of generality, the optimal measurement that discriminates the states $\rho_{n,\sigma}$ can be represented by a POVM whose elements have the form shown in Eq.~\eqref{app:rho_block}. Moreover, we can assume it to be covariant under $S_N$~\cite{Holevo1982}.
So, such POVM elements can be written as
\begin{equation}\label{app:povmelements}
E_{n,\sigma} = \bigoplus_\l \id_\la\otimes  U^\l_\sigma\, \Xi_\lb^n (U^\l_\sigma)^\dagger \,,
\end{equation}
where $\Xi_\lb^n$ is some positive operator. The resolution of the identity condition imposes constraints on them. 
The condition reads
\begin{align}
\begin{split}\,
\sum_{n,\sigma} E_{n,\sigma} &= \sum_n \frac{1}{b_n} \sum_{\sigma\in S_N} \bigoplus_\l \id_\la\otimes U^\l_\sigma \Xi_\lb^n (U^\l_\sigma)^\dagger \\
&= \bigoplus_\l \id_\la\otimes\id_\lb \,,
\end{split}
\end{align}
where we have used the factor $b_n$ to extend the sum over~${\mathscr S}_n$ to the entire group $S_N$ and applied Schur lemma. Taking the trace on both sides of the equation, we find the POVM constraint to be
\begin{equation}\label{povmcond}
\sum_n \frac{N!}{b_n} \tr{\left(\Xi_\lb^n\right)} = \nu_\l \,,\quad \forall \l \,,
\end{equation}
where $\nu_\l$ is the dimension of $\openone_\lb$ or, equivalently, the multiplicity of the irrep $\l$ of ${\rm SU}(d)$ [see Eq.~(\ref{mult_nu})]. 

So far we have analyzed the structure that the symmetries of the problem impose on the states $\rho_{n,\sigma}$ and the measurements. We have learned that for any choice of  operators $\Xi^n_{\lb}$ that fulfill Eq.~(\ref{povmcond}), the set of operators~(\ref{app:povmelements}) defines a valid POVM, but it need not be optimal. So, we now proceed to derive optimality conditions for $\Xi^n_{\lb}$. Those are provided by the Holevo-Yuen-Kennedy-Lax~\cite{Holevo1973a,Yuen1975} necessary and sufficient conditions for minimizing the average cost. For our clustering problem in Eq.~(\ref{app:av_cost}) they read 
\begin{align}
\label{Holevo1}
	&(W_{\x}-\Gamma)E_\x=E_\x(W_\x-\Gamma)=0 \,,\\
\label{Holevo2}
	&\phantom{(}W_{\x}-\Gamma \geq 0 \,.
\end{align}
They must hold for all $\x$, where $\Gamma=\sum_{\x} W_\x E_\x=\sum_\x E_\x W_\x$, and $W_\x = \sum_{\x'} f(\x,\x') \eta_{\x'} \rho_{\x'}$.
We will assume that the prior distribution $\eta_{\x}$ is flat and that the cost function is nonnegative and covariant with respect to the permutation group, i.e., $f(\x,\x')=f(\tau\x,\tau\x')$ for all $\tau\in S_N$. Then, $W_{\tau\x}=U_\tau W_{\x} U_\tau^\dagger$ 
and we only need to ensure that conditions \eqref{Holevo1} and \eqref{Holevo2} are met for reference strings, for which $\x=(n,e)$.
In the Schur basis, their corresponding  operators, which we simply call $W_n$, and the matrix $\Gamma$ take the form
\begin{align}
W_n &= \bigoplus_\l \id_\la \otimes \omega^n_\lb \,,\label{W_block}\\
\Gamma &= \bigoplus_\l k_\l \id_\la\otimes\id_\lb\,,\label{G_block}
\end{align}
where we have used Schur lemma to obtain Eq.~(\ref{G_block}) and defined $k_\l \equiv \sum_n N!\, \tr{\left(\omega^n_\lb \,\Xi^n_\lb\right)}/(b_n\nu_\l)$. 
Note that $\Gamma$ is a diagonal matrix, in spite of the fact that $\omega_\lb^n$ are, at this point, arbitrary full-rank positive operators.

With Eqs.~\eqref{W_block} and \eqref{G_block}, the optimality conditions~\eqref{Holevo1} and \eqref{Holevo2} can be made explicit. First, we note that the subspace $\la$ is irrelevant in this calculation, and that there will be an independent condition for each irrep $\l$. Taking into account these considerations, Eq.~\eqref{Holevo1} now reads
\begin{align}
\omega^n_\lb\Xi^n_\lb &=\Xi^n_\lb\omega^n_\lb =k_\l \Xi^n_\lb \,,\quad \forall n,\l \,.\label{Holevo1b}
\end{align}
This equation tells us two things: (i) since the matrices~$\omega^n_\lb$ and~$\Xi^n_\lb$ commute, they have a common eigenbasis, and (ii)~Eq.~\eqref{Holevo1b} is a set of eigenvalue equations for $\omega^n_\lb$ with a common eigenvalue $k_\l$, one equation for each eigenvector of $\Xi^n_\lb$. Therefore, the support of $\Xi^n_\lb$ is necessarily restricted to a single eigenspace of $\omega^n_\lb$. Denoting by $\vartheta_{\l,a}^n$, $a=1,2,\dots$, the eigenvalues of $\omega_\lb^n$ sorted in increasing order, we have $k_\l=\vartheta_{\l,a}^n$ for some $a$, which may depend on $\lambda$ and $n$, or else $\Xi^n_\lb=0$.

The second Holevo condition~\eqref{Holevo2}, under the same considerations regarding the block-diagonal structure, leads to
\begin{equation}\label{Holevo2b}
\omega_\lb^n \geq k_\l \id_\lb \,,\quad \forall n,\l \,.
\end{equation}
This condition further induces more structure in the POVM. 
Given $\l$, Eq.~\eqref{Holevo2b} has to hold for {\em every} value of~$n$. In particular, 
we must have $\min_{n'} \vartheta_{\l,1}^{n'}\ge k_\l$. 
Therefore, $\min_{n'} \vartheta_{\l,1}^{n'}\ge\vartheta^n_{\l,a}$ for some $a$, or else $\Xi^n_\lb=0$. Since~$\Xi^n_\lb$ cannot vanish for all $n$ because of Eq.~(\ref{povmcond}),  we readily see that 
\begin{equation}\label{povm}
k_\l\!=\!\vartheta^{n(\l)}_{\l,1}, \quad
\Xi_\lb^n \!= 
\begin{cases}
\xi_\l^n \Pi_{1}(\omega_\lb^n) & {\rm if} \, n=n(\l), \\
0 & {\rm otherwise},
\end{cases}
\end{equation}
where $n(\l)={\rm argmin}_n \vartheta_{\l,1}^n$, $\Pi_{1}(\omega_\lb^n)$ is a projector onto the eigenspace of $\omega_\lb^n$ (not necessarily the whole subspace) corresponding to the minimum eigenvalue $\vartheta_{\l,1}^n$, and $\xi^n_\lambda$ is a suitable coefficient that can be read off from~Eq.~\eqref{povmcond}:
\begin{equation}\label{povmcoef}
\xi_\l^{n} = \frac{\nu_\l b_n}{D_\l^{n} N!}  \,,
\end{equation}
where $D_\l^n = \dim{[\Pi_{1}(\omega_\lb^n)]}$. This completes the construction of the optimal POVM.

For a generic cost function, we can now write down a closed, implicit formula for the minimum average cost achievable by any quantum clustering protocol. It reads
\begin{equation}\label{opt_av_cost}
\bar f = \tr \Gamma = \sum_\l s_\l \,\nu_\l\, \vartheta_{\l,1}^{n(\l)}\,,
\end{equation}
where $s_\l$ is the dimension of $\openone_\la$ or, equivalently, the multiplicity of the irrep $\l$ of $S_N$ [see Eq.~(\ref{mult_s})]. The only object that remains to be specified is the function~$n(\l)$, which depends ultimately on the choice of the cost function $f(\x,\x')$.

\subsubsection{Success probability}

We now make Eq.~\eqref{opt_av_cost} explicit by considering the success probability $P_{\rm s}$ as a figure of merit, that is, we choose $f(\x,\x')=1-\delta_{\x,\x'}$, hence $P_{\rm s}=1-\bar f$. We also assume that the source that produces the input sequence is equally likely to prepare either state, thus each string~$\x$ has the same prior probability, $\eta_{\x} = 2^{1-N} \equiv\eta$. In this case,~$W_n$ takes the simple form
\begin{equation}\label{Wme}
W_n = 
\bigoplus_\l \id_\la \otimes \left(\mu_\l\id_\lb -\eta c_n \Omega_\lb^n\right) \,,
\end{equation}
where $\mu_\lambda$ are positive coefficients and we recall that the  expression in parenthesis corresponds to $\omega_\lb^n$ in Eq.~\eqref{W_block}. From this expression one can easily derive the explicit forms of $\vartheta_{\l,1}^{n}$ and $n(\l)$. We just need to consider the maximum eigenvalue of the rank-one projector~$\Omega^n_\lb$, which can be either one or zero depending on whether  or not the input state $\rho_{n,\sigma}$ has support in the irrep $\l$ space. So, among the values of $n$ for which $\rho_{n,\sigma}$ does have support there, $n(\l)$ is one that maximizes $c_n$. Since $c_n$ is a decreasing function of $n$ in its allowed range (recall that $n\le\floor{N/2}$), $n(\l)$ is the smallest such value.

For the problem at hand, the irreps in the direct sum can be labeled by Young diagrams of at most two rows, or, equivalently, by partitions of $N$ of length at most two  (see Appendix~\ref{app:irreps}), hence $\l=(\l_1,\l_2)$, where $\l_1+\l_2=N$ and $\l_2$ runs from $0$ to $\floor{N/2}$.
Given~$\l$, only states~$\rho_n$ with $n=\l_2,\ldots,\floor{N/2}$ have support on the irrep $\lambda$ space, as readily follows from the Clebsch-Gordan decomposition rules.
Then, 
\begin{equation}\label{nl}
n(\l) =  \l_2\,,\quad
\vartheta^{n(\l)}_{\l,1}=\mu_\l-\eta c_{n(\l)} \,.
\end{equation}
Eq.~\eqref{nl} gives the optimal guess for the size, $n$, of the smallest cluster. The rule is in agreement with our intuition. The irrep $(N,0)$, i.e., $\l_2=0$, corresponding to the fully symmetric subspace, is naturally associated with the value $n=0$, i.e., with all $N$ systems being in the same state/cluster; the irrep with one antisymmetrized index has $\l_2=1$, and hints at a system being in a different state than the others, i.e., at a cluster of size one; and so on.  

We now have all the ingredients to compute the optimal success probability from~Eq.~(\ref{opt_av_cost}). It reads
\begin{align}
P_{\rm s} 
&= \eta \sum_\l c_{n(\l)} s_\l \nu_\l \nonumber\\
&= {1\over2^{N-1}}\sum_{i=0}^{\floor{N/2}} \binom{N}{i} \frac{(d-1)(N-2i+1)^2}{(d+i-1)(N-i+1)^2} \,, \label{app:ps}
\end{align}
where we have used the relation
$\sum_\l s_\l \nu_\l \mu_\l=1$ that follows from  $\tr \sum_\x \eta_\x \rho_\x=1$, and
the expressions of $\nu_\l$ and~$s_\l$ from Eqs.~\eqref{mult_nu} and \eqref{mult_s} in Appendix~\ref{app:irreps}.

\subsection{Clustering quantum states: known input states}\label{sec:quantumknown}

If the two possible states $\ket{\phi_0}$ and $\ket{\phi_1}$ are known, the optimal clustering protocol must use this information. It is then expected that the average performance will be much higher than for the universal protocol. It is natural in this context not to identify a given string~$\x$  with its complementary $\bar\x$  (we stick to the notation in the main text), since mistaking one state for the other should clearly count as an error if the two preparations are specified. In this case, then, clustering is equivalent to discriminating the $2^N$ known pure states $\ket{\Phi_\x}\!=\!\ket{\phi_{x_1}\rangle\!\otimes\!|\phi_{x_2}\rangle\!\otimes\!\cdots\! \otimes\!|\phi_{x_N}}$ (hypotheses), 
where with no loss of generality we can write
\begin{equation}
\ket{\phi_{0/1}}=\sqrt{\frac{1+c}{2}}\ket{0}\pm \sqrt{\frac{1-c}{2}}\ket{1}
\label{no loss}
\end{equation}
for a convenient choice of basis. Here $c=|\langle\phi_0|\phi_1\rangle|$ is the overlap of the two states.

The Gram matrix $G$ 
encapsulates all the information needed to discriminate the states of the set. It is defined as having elements 
$G_{\x,\x'}=\braket{\Phi_\x}{\Phi_{\x'}}$.  It is known that when the diagonal elements of  its square root are all equal, i.e., $\big(\sqrt{G}\,\big)_{\x,\x}\equiv S$ for all $\x$,  
then the square root measurement is optimal~\cite{DallaPozza2015,Sentis2016} and the probability of successful indentification reads
simply $P_{\rm s}=S^2$. Notice that we have implicitly assumed uniformly distributed hypotheses.
For the case at hand,
\begin{align}
G_{\x,\x'}&=(\langle\phi_{x_1}|\otimes\cdots\otimes\langle\phi_{x_N}|)(|\phi_{x'_1}\rangle\otimes\cdots\otimes|\phi_{x'_N}\rangle)\nonumber
\\
&=\prod_{i=1}^N\langle\phi_{x_i}|\phi_{x'_i}\rangle
=\left({\mathscr G}^{\otimes N}\right)_{\x,\x'},
\end{align}
where
\begin{equation}
{\mathscr G}=\begin{pmatrix} 1 & c \\ c & 1 \end{pmatrix} 
\end{equation}
is the Gram matrix of $\{|\phi_0\rangle,|\phi_1\rangle\}$. 
Thus, $\sqrt G\!=\!(\sqrt{\mathscr G}\,)^{\otimes N}$, with
\begin{equation}
\sqrt{{\mathscr G}}=\begin{pmatrix}  \displaystyle \frac{\sqrt{1\!+\!c}+\!\sqrt{1\!-\!c}}{2} & \displaystyle   \frac{\sqrt{1\!+\!c}-\!\sqrt{1\!-\!c}}{2} \\[.8em] 
  \displaystyle \frac{\sqrt{1\!+\!c}-\!\sqrt{1\!-\!c}}{2} &   \displaystyle\frac{ \sqrt{1\!+\!c}+\!\sqrt{1\!-\!c}}{2} \end{pmatrix}.
\end{equation}
As expected, the diagonal terms of $\sqrt{G}$ are all equal, 
and the success probability is given by
\begin{equation}\label{psnc}
P_{\rm s}(c)\!=\!\left( \!\sqrt{1\!+\!c}\!+\!\sqrt{1\!-\!c}\over2\, \right)^{\!2N}\!\!\!=\left( 1\!+\!\sqrt{1\!-\!c^2}\over2\, \right)^{\!N}.
\end{equation}
We call the reader's attention to the fact that one could have attained the very same success probability by performing an individual Helstrom measurement~\cite{Helstrom1976}, with basis
\begin{equation}
\ket{\psi_{0/1}}=\frac{\ket{0}\pm \ket{1}}{\sqrt{2}},
\label{Helstrom basis}
\end{equation}
 on each state of the input sequence and guessed that the label of that state was the outcome value.
In other words, for the problem at hand, global quantum measurements do not provide any improvement over individual fixed measurements.

In order to compare with the results of the main text, we compute the average performance for a uniform distribution of states $\ket{\phi_0}$ and $\ket{\phi_1}$, i.e., the average
\begin{align}
P_{\rm s}\! &=\!\!\int\! d\phi_0 d\phi_1 P_{\rm s}(c)\nonumber\\
&=\!\!\int_0^1\! dc^2 P_{\rm s}(c)\!\int \!d\phi_0 d\phi_1\,\delta\!\left(|\langle\phi_0|\phi_1\rangle|^2\!-\!c^2\right)\nonumber\\
&=\!\!\int_0^1\! dc^2 P_{\rm s}(c)\!\int \!d\phi_1\,\delta\!\left(|\langle 0|\phi_1\rangle|^2\!-\!c^2\right)\nonumber\\
&=\!\!\int_0^1\! dc^2 \mu(c^2)P_{\rm s}(c),
\end{align}
where we have inserted the identity $1=\int_0^1dc^2 \delta(a^2-c^2)$, for $0<a\equiv |\langle\phi_0|\phi_1\rangle| < 1$, and used the invariance of the measure~$d\phi$ under ${\rm SU}(d)$ transformations. 
The marginal distribution is $\mu(c^2)=(d-1) (1-c^2)^{d-2}$ (see Appendix~\ref{app:prior}).
Using this result, the asymptotic behavior of the last integral is 
\begin{equation}\label{eq:meanqudits}
P_{\rm s}\sim \frac{4(d-1)}{N}\,.
\end{equation}
As expected, knowing the two possible states in the input string leads to a better behavior of  the success probability: it decreases only linearly in $1/N$, as compared to the best universal quantum clustering protocol, which exhibits a quadratic decrease.

To do a fairer comparison with universal quantum clustering, guessing the complementary string $\bar\x$ instead of $\x$ will now be counted as success, that is,
now the clusterings are defined by the states
\begin{equation}
\rho_\x=\frac{\ketbra{\Phi_\x}{\Phi_\x}+
 \ketbra{\Phi_{\bar{\x}}}{\Phi_{\bar{\x}}}}{2}.
\end{equation}
For this variation of the problem, the optimal measurement is still local, and given by a POVM with elements
\begin{equation}
E_\x=\ketbra{\Psi_\x}{\Psi_\x}+\ketbra{\Psi_{\bar{\x}}}{\Psi_{\bar{\x}}},
\end{equation}
where $\ket{\Psi_\x}\!=\!\ket{\psi_{x_1}\rangle\!\otimes\!|\psi_{x_2}\rangle\!\otimes\!\cdots\! \otimes\!|\psi_{x_N}}$, and where we recall that $\{\ket{\psi_{0}},\ket{\psi_{1}}\}$ is the (local) Helstrom measurement basis in Eq.~(\ref{Helstrom basis}). Note that $\{E_\x\}$ are orthogonal projectors.

To prove the statement in the last paragraph, we show that the Holevo-Yuen-Kennedy-Lax conditions, Eq.~(\ref{Holevo1}),  hold (recall that the Gram matrix technique does not apply to mixed states). For the success probability and assuming equal priors, these conditions take the simpler form
\begin{align}
\sum_\x E_\x \rho_\x&=\sum_\x \rho_\x E_\x\equiv\Gamma,\label{holevo-cond0}\\
\Gamma-\rho_\x&\geq 0 \quad \forall \x,
\label{holevo-cond}
\end{align}
where we have dropped the irrelevant factor $\eta=2^{1-N}$.
Condition~(\ref{holevo-cond0}) is trivially satisfied. To check that condition~(\ref{holevo-cond}) also holds,
we recall the Weyl inequalities for the eigenvalues of Hermitian $n\times n$ matrices $A$, $B$~\cite{Horn2013}:
\begin{equation}
\vartheta_i(A+B)\leq \vartheta_{i+j}(A)+\vartheta_{n-j}(B),
\label{Weyl ineq}
\end{equation}
for $j=0,1,\ldots,n-i$,
where the eigenvalues are labeled in increasing order $\vartheta_1\leq\vartheta_2\leq\cdots \leq\vartheta_n$. We use Eq.~(\ref{Weyl ineq}) to write 
 \begin{equation}
\vartheta_1(\Gamma)\leq \vartheta_{3}(\Gamma-\rho_\x)+\vartheta_{2^{N}-2}(\rho_\x)
\label{rmx}
\end{equation}
(note that effectively all these operators act on the $2^N$-dimensional subspace spanned by $\{|0\rangle,|1\rangle\}^{\otimes N}$).
As will be proved below,  $\Gamma>0$, which implies that $\vartheta_1(\Gamma)>0$. We note that $\rho_\x$ has rank two, i.e., it has only two strictly positive eigenvalues, so
$\vartheta_{2^N-2}(\rho_\x)=0$. Then Eq.~(\ref{rmx}) implies 
  \begin{equation}
  \label{lambda3}
 \vartheta_{3}(\Gamma-\rho_\x) \geq \vartheta_1(\Gamma)>0.
\end{equation}
Finally,
notice that $\Gamma-\rho_\x$ has two null eigenvalues, with eigenvectors $\ket{\Psi_\x}$ and $\ket{\Psi_{\bar{\x}}}$. Hence,   
 $\vartheta_1(\Gamma-\rho_\x)=\vartheta_2(\Gamma-\rho_\x)=0$, and it follows from Eq.~\eqref{lambda3}  that  condition~(\ref{holevo-cond}) must hold.
 
To show the positivity of $\Gamma$, which was assumed in the previous paragraph, we use Eqs.~(\ref{no loss}) and~(\ref{Helstrom basis}) to write
\begin{align}
\Gamma =\frac{1}{2}\left[ \begin{pmatrix} a_1 & 0 \\ 0 &a_2 \end{pmatrix}^{\!\!\otimes N} 
                \!\!\! +\begin{pmatrix} b_1& 0 \\ 0 & b_2 \end{pmatrix}^{\!\!\otimes N} \right],
\end{align}
where 
\begin{align}
a_{1/2}=&\frac{1\pm c+\sqrt{1-c^2} }{2},\nonumber \\
b_{1/2}=&\frac{ 1\pm c - \sqrt{1-c^2} }{2}.
\end{align}
Notice that 
$a_1>b_1$ and $a_2>|b_2|$. Thus, if $0\leq c< 1$, we have
 $\vartheta_k >0$ for $k=1,2,\dots, 2^{N}$.
The special case~$c=1$ is degenerate. Eq.~\eqref{holevo-cond} is trivially saturated, rendering $P_{\rm s}=2^{1-N}$, as it should be.

The maximum success probability can now be computed recalling that $P_{\rm s}(c)=2^{1-N} \tr\Gamma$. We obtain
\begin{equation}
P_{\rm s}(c)=\left(\frac{1+\sqrt{1-c^2}}{2}\right)^{\!\!N}\!\!+\left(\frac{1-\sqrt{1-c^2}}{2}\right)^{\!\!N},
\end{equation}
where the first term corresponds to guessing correctly all the states in the input string, whereas the second one results from guessing the other possible state all along the string.
One can easily check that the average over~$c$ of the second term vanishes exponentially for large $N$, and we end up with a success probability given again by~Eq.~\eqref{eq:meanqudits}.

Finally, we would like to mention that one could consider a simple unambiguous protocol~\cite{Ivanovic1987,Dieks1988,Peres1988,Chefles1998a} whereby each state of the input string would be identified with no error with probability $P_{\rm s}(c)=1-c$, i.e., the protocol would give an inconclusive answer with probability $1-P_{\rm s}=c$. Therefore, the average unambiguous probability of sorting the data would be
\begin{equation}
P_\mathrm{s}=2\!\! \int_0^1\!\! dc\,c \mu(c^2)(1-c)^N \sim \frac{2(d-1)}{N^2}\,.
\end{equation}
}

\section{Discussion}\label{sec:discussion}

Unsupervised learning, which assumes virtually nothing about the distributions underlying the data, is already a hard problem~\cite{Aloise2009,Ben-David2015}. Lifting the notion of classical data to quantum data (i.e., states) factors in additional obstacles, such as the impossibility to repeatedly operate with the quantum data without degrading it. 
\blue{Most prominent classical clustering algorithms heavily rely on the iterative evaluation of a function on the input data (e.g., pairwise distances between points in a feature vector space, as in $k$-means~\cite{Lloyd1982}), hence they are not equipped to deal with degrading data and would expectedly fail in our scenario.
The unsupervised quantum classification algorithm we present is thus, by necessity, far away from its classical analogues. 
In particular, since we are concerned with the optimal quantum strategy we need to consider the most general collective measurement, which is inherently single-shot:} 
it yields a single sample of a stochastic action, namely, a posterior state and an outcome of a quantum measurement, where the latter provides the description of the clustering.
The main lesson stemming from our investigation is that, despite these limitations, clustering unknown quantum states is a feasible task. 
\blue{The optimal protocol that solves it showcases some interesting features.}

\blue{
{\em It does not completely erase the information about a given preparation of the input data after clustering.}}
This is apparent from Eq.~\eqref{povm_elements_main}, since the action of the POVM on the subspaces~$\la$ is the identity.
\blue{After the input data string in the global state $\ket{\Phi_\x}$ is measured and outcome $\l^*$ is obtained (recall that $\l^*$ gives us information about the size of the clusters), information relative to the particular states $\ket{\phi_{0/1}}$ remains in the subspace~$(\l^*)$ of the global post-measured state.
Therefore, one could potentially use further the posterior (clustered) states down the line as approximations of the two classes of states. This opens the door for our clustering device to be used as an intermediate processor in a quantum network.}
This notwithstanding, the amount of information that can be retrieved after optimal clustering is currently under investigation. 

{\em It outbeats the classical and semiclassical protocols.} If the local dimension of the quantum data is larger than two, the dimensionality of the symmetric subspaces spanned by the global states of the strings of data can be exploited by means of collective measurements with a twofold effect: enhanced distinguishability of states, resulting in improved clustering performance (exemplified by a linear  increase in the asymptotic success probability), and information-preserving data handling (to some extent, as discussed above). This should be contrasted with the semiclassical protocol, which essentially obliterates the information content of the data (as a von Neumann measurement is performed on each system), and whose success probability vanishes exponentially with the local dimension.  In addition, the optimal classical and semiclassical protocols require solving an NP-complete problem and their implementation is thus inefficient. In contrast, we observe that the first part of the quantum protocol, which consists in guessing the size of the clusters $n$, runs efficiently on a quantum computer: this step involves a Schur transform that runs in polynomial time in $N$ and $\log d$~\cite{Harrow2005,Krovi2018}, followed by a projective measurement with no computational cost. The second part, guessing the permutation $\sigma$, requires implementing a group-covariant POVM. The complexity of this step, and hence the overall computational complexity of our protocol, is still an open question currently under investigation.

{\em It is optimal for a range of different cost functions.} There are various cost functions that could arguably be better suited to quantum clustering, e.g., the Hamming distance between the guessed and the true clusterings, or likewise, the trace distance or the infidelity between the corresponding effective states~$\rho_{n,\sigma}$ and~$\rho_{n',\sigma'}$. They are however hard to deal with analytically. The question arises as to whether our POVM is still optimal for such cost functions. To answer this question, we formulate an optimality condition that can be checked numerically for problems of finite size (see Appendix~\ref{app:generalcosts}). Our numerics show that the POVM remains optimal for all these examples. This is an indication that the optimality of our protocol stems from the structure of the problem, independently of the cost function. 
 
{\em It stands a landmark in multi-hypothesis state discrimination.}  
Analytical solutions to multi-hypothesis state discrimination exist only in a few specific cases~\cite{Barnett2001,Chiribella2004,Chiribella2006a,Krovi2015a,Sentis2016,Sentis2017}. Our set of hypotheses arises arguably from the minimal set of assumptions about a pure state source: it produces two states randomly. 
Variants of this problem with much more restrictive assumptions 
have been considered in Refs.~\cite{Korff2004,Hillery2011,Skotiniotis2018}.

Our clustering protocol departs from other notions of quantum unsupervised machine learning that can be found in the literature~\cite{Aimeur2013,Lloyd2013,Wiebe2014a,Kerenidis2018}. In these references, data coming from a classical problem is encoded in quantum states that are available on demand via a quantum random access memory~\cite{Giovannetti2008}. The goal is to surpass classical performance in the number of required operations. In contrast, we deal with unprocessed quantum data as input, and aim at performing a task that is genuinely quantum. This is a notably harder scenario, where known heuristics for classical algorithms simply cannot work. 

%

\blue{Other extensions of this work currently under investigation are: clustering systems whose states can be of more than two types, where we expect a similar two-step measurement for the optimal protocol; and clustering of quantum processes, where the aim is to classify instances of unknown processes by letting them run on some input test state of our choice (see Ref.~\cite{Skotiniotis2018} for related work on identifying malfunctioning devices).
In this last case, an interesting application arises when considering causal relations as the defining feature of a cluster. A clustering algorithm would then aim to identify, within a set of unknown processes, which ones are causally connected. Identifying causal structures has recently attracted attention among the quantum information community~\cite{Chiribella2019}.}

\acknowledgments
We acknowledge the financial support of the Spanish MINECO, ref. FIS2016-80681-P (AEI/FEDER, UE), and Generalitat de Catalunya CIRIT, ref. 2017-SGR-1127. GS thanks the support of the Alexander von Humboldt Foundation.
EB also thanks the hospitality of Computer Science Department of the University of Hong Kong during his stay.

\appendix

\section{Partitions}\label{app:partitions}

Partitions play an important role in the representation theory of groups and are central objects in combinatorics. Here, we collect a few definitions and  results that are used in the next appendices, particularly in Appendix~\ref{app:irreps}.

A \emph{partition} $\l=(\l_1,\l_2,\ldots,\l_r,\ldots)$ is a sequence of nonnegative integers in nonincreasing order. 
The \emph{length}~of~$\l$, denoted $l(\l)$, is the number of nonzero elements in~$\l$.
We denote by \mbox{$\l\vdash N$} a partition $\l$ of the integer $N$, where $N=\sum_i \l_i$. 
A natural way of ordering partitions is by inverse lexicographic order, i.e.,  given two partitions $\l$ and $\l'$, we write $\l > \l' $ iff the first nonzero difference $\l_i-\l'_i$ is positive. 

The total number of partitions of an integer $N$ is denoted by~$P_N$~\cite{Flajolet2009}, and the number of partitions such that $l(\l)\le r$ by~$P^{(\le r)}_N$. Similarly, the number of partitions of length~$r$ is denoted by~$P^{(r)}_N$. There exists no closed expression for any of these numbers, but there are widely known results (some of them by Hardy and Ramanujan are very famous~\cite{Andrews1976}) concerning their asymptotic behavior for large $N$. The one we will later use in Appendix~\ref{app:classic} is
\begin{equation}
P^{(\le r)}_N\sim {N^{r-1}\over r!(r-1)!}\,,
\label{partAsym}
\end{equation}
which gives the dominant contribution for large $N$. Note that from the obvious relation $P^{(r)}_N=P^{(\le r)}_N-P^{(\le r-1)}_N$, it follows that the same asymptotic expression holds for~$P^{(r)}_N$.

Partitions are conveniently represented by \emph{Young diagrams}. The Young diagram associated to the partition $\l \vdash N$
is an arrangement of $N$ empty boxes in $l(\l)$ rows, with~$\l_i$ boxes in the $i$th row. This association is one-to-one, hence $\l$ can be used to label Young diagrams as well. A \emph{Young tableau} of $d$ entries is a Young diagram filled with integers from 1 up to~$d$, one in each box. There are two types of tableaux: A \emph{standard Young tableau} (SYT) of shape $\l \vdash N$ is one where $d=N$ and such that the integers in each row increase from left to right, and  from top to bottom in each column (hence each integer appears exactly once). A \emph{semistandard Young tableau} (SSYT) of shape $\l \vdash N$ and $d$ entries,  $d\geq l(\l)$, is one such that integers in each row are nondecreasing from left to right, and increasing from top to bottom in each column.

The number of different SYTs of shape $\l \vdash N$ is given by the \emph{hook-length} formula
\begin{equation}\label{mult_nu_general}
\nu_\l = \frac{N!}{\prod_{(i,j)\in\l} h_{ij}} \,,
\end{equation}
where $(i,j)$ denotes the box located in the $i$th row and the $j$th column of the Young diagram, and $h_{ij}$ is the hook-length of the box~$(i,j)$, defined as the number of boxes 
located beneath or to the right of that box in the Young diagram, counting the box itself.
Likewise, the number of SSYTs of shape $\l \vdash N$ and $d$ entries is given by the formula
\begin{equation}\label{mult_s_general}
s_\l = \frac{\Delta(\l_1+d-1,\l_2+d-2,\ldots,\l_d)}{\Delta(d-1,d-2,\ldots,0)} \,,
\end{equation}
where $\Delta(x_1,x_2,\dots,x_d)=\prod_{i<j}(x_i-x_j)$.

\section{Irreducible representations of SU$(d)$ and $S_N$ over $(d,\mathbb{C})^{\otimes N}$}\label{app:irreps}

For the sake of convenience, we recall here some ingredients of representation theory that we use throughout the paper. 
The results described below can be found in standard textbooks, for instance, in Refs. \cite{Sagan2001,Goodman2009}.

\subsection{Some results in representation theory}

Young diagrams or, equivalently, partitions $\l$, label the irreducible representations (irreps) of the general linear group GL$(d)$ and some of its subgroups, e.g., ${\rm SU}(d)$, and also the irreps of the symmetric group $S_N$. The dimension of these irreps are given by $s_\l$ and $\nu_\l$, respectively [Eqs.~(\ref{mult_nu_general}) and~(\ref{mult_s_general})].

Schur-Weyl duality \cite{Goodman2009} establishes a connection between irreps of both groups, as follows. Let us consider the transformations $R^{\otimes N}$ and $U_\sigma$ on the $N$-fold tensor product space $(d,\mathbb{C})^{\otimes N}\!$, where $R\in {\rm SU}(d)$ and $U_\sigma$ permutes the $N$ spaces $(d,\mathbb{C})$ of the tensor product according to the permutation $\sigma\in S_N$. Both $R^{\otimes N}$ and $U_\sigma$ define, respectively, a reducible unitary representation of the groups SU$(d)$ and~$S_N$ on $(d,\mathbb{C})^{\otimes N}\!$. Moreover, they are each other's commutants. 
It follows that this reducible representation decomposes into irreps $\l$, so that their joint action can be expressed as
\begin{equation}\label{schurweyl}
R^{\otimes N}U_\sigma= U_\sigma R^{\otimes N} = \bigoplus_{\l \vdash N} R^\l \otimes U^\l_\sigma \,,
\end{equation}
where $R^\l$ and $U^\l_\sigma$ are the matrices that represent $R$ and $U_\sigma$, respectively, on the irrep~$\l$. To resolve any ambiguity that may arise, we write $\l$ in parenthesis, $\la$, when it refers to the irreps of SU$(d)$, or in brackets,~$\lb$, when it refers to those of~$S_N$.   Eq.~(\ref{schurweyl}) tells us that the dimension of $\la$, $s_\l$, coincides with the \emph{multiplicity} of $\lb$, and conversely, the dimension of $\lb$, $\nu_\l$, coincides with the multiplicity of $\la$. 

This block-diagonal structure provides a decomposition of Hilbert space $\mathcal{H}^{\otimes N}=(d,{\mathbb C})^{\otimes N}$ into subspaces that are invariant under the action of ${\rm SU}(d)$ and $S_N$, as $\mathcal{H}^{\otimes N}=\bigoplus_\l H_\l$, and in turn, $H_\l = H_\la \otimes H_\lb$.
The basis in which ${\mathcal H}^{\otimes N}$ has this form is known as \emph{Schur basis}, and the unitary transformation that changes from the computational to the Schur basis is called \emph{Schur transform}.

To conclude this Appendix, let us recall the rules for reducing the tensor product of two ${\rm SU}(d)$ representations as a Clebsch-Gordan series of the form
\begin{equation}\label{reduction1}
R^\l\otimes R^{\l'}=\bigoplus_{\l''} R^{\l''}\otimes \openone^{\lambda''} \,,\quad \forall R\in{\rm SU}(d)\,,
\end{equation}
where ${\rm dim}(\id^{\lambda''})$ is the multiplicity of irrep $\lambda''$. 
The same rules also apply to the reduction of the outer product of representations of $S_n$ and~$S_{n'}$ into irreps of~$S_{n''}$, where $n''=n+n'$. In this case one has
\begin{equation}\label{reduction2}
(U^\l \otimes U^{\l'})_{\sigma}=\bigoplus_{\l''} U^{\l''}_{\sigma} \otimes \openone^{\lambda''}, \quad \forall \sigma\in S_{n''}.
\end{equation}
Note the different meanings of $\otimes$ in the last two equations (it is however standard notation). The rules are most easily stated in terms of the Young diagrams  that label the irreps. They are as follows:
\begin{enumerate}
\item In one of the diagrams that label de irreps on the left hand side of Eq.~(\ref{reduction1}) or Eq.~(\ref{reduction2}) (preferably the smallest), write the symbol $a$ in all boxes of the first row, the symbol $b$ in all boxes of the second row, $c$ in all boxes of the third one, and so~on.
\item Attach boxes with $a$ to the second Young diagram in all possible ways subjected to the rules that no two $a$'s appear in the same column and that the resulting arrangement of boxes is still a Young diagram. Repeat this process with $b$'s, $c$'s, and so~on.
\item For each Young diagram obtained in step two, read the 1st row of added symbols from right to left, then the second row in the same order, and so on. The resulting sequence of symbols, e.g., $abaabc\dots$, must be a lattice permutation, namely, to the left of any point in the sequence, there are not fewer $a$'s than $b$'s, no fewer $b$'s than $c$'s, and so on. Discard all diagrams that do not comply with this rule.
\end{enumerate}

The Young diagrams $\l''$ that result from this procedure specify the irreps on the right hand side of Eqs.~(\ref{reduction1}) and~(\ref{reduction2}). A same diagram can appear a number $M$ of times, in which case $\l''$ has multiplicity ${\rm dim}(\openone^{\lambda''})=M$.

\subsection{Particularities of quantum clustering}

Since the density operators [cf. Eq.~\eqref{app:rho_ns}] and POVM elements [cf. Eq.~\eqref{app:povmelements}] associated to each possible clustering emerge from the joint action of a permutation $\sigma\in S_N$ and a group average over ${\rm SU}(d)$, it is most convenient to work in the Schur basis, where the mathematical structure is much simpler. A further simplification specific to quantum clustering of two types of states, is that the irreps that appear in the block-diagonal decomposition of the states (and, hence, of the POVM elements) have at most length 2, i.e., they are labeled by bipartitions $\l=(\l_1,\l_2)$, and correspond to Young diagrams of at most two rows. This is because the $\rho_{n,\sigma}$ arise from the tensor product of two \emph{completely symmetric} projectors, $\openone^{\sym}_n$,  $\openone^{\sym}_{\bar n}$, of~$n$ and~$\bar{n}$ systems [cf. Eq.~\eqref{app:rho_ns}]. They project into the irrep $\l=(n,0)$ and $\l'=(\bar n,0)$ subspaces, respectively. 
According to the reduction rules above, in the Schur basis the tensor product reduces as
\ytableausetup{mathmode,boxsize=1em}
\ytableausetup{centertableaux}
\begin{eqnarray}
&&
\overbrace{
\begin{ytableau}
  \phantom{.}  & \phantom{.} 
 \end{ytableau}
\cdots
 \begin{ytableau}
    \phantom{.}  &  \phantom{.} &  \phantom{.}  
\end{ytableau}
}^{\bar n}
 \ \otimes \
\overbrace{ \begin{ytableau}
a & a
 \end{ytableau}
\cdots
 \begin{ytableau}
  a &  a 
\end{ytableau}}^{n}\nonumber\\
&=&
\overbrace{
\begin{ytableau}
 \phantom{.} & \phantom{.} & \phantom{.}
 \end{ytableau}
\cdots
 \begin{ytableau}
   a  &  a &  a
\end{ytableau}
}^{n+\bar n} 
\ \oplus\
\overbrace{
\begin{ytableau}
 \phantom{.} & \phantom{.} & \phantom{.}\\
 a
 \end{ytableau}
 \raisebox{.5em}{
$\cdots$\!
 \begin{ytableau}
   a  &  a
\end{ytableau}
}\!\!
}^{n+\bar n-1}
\label{reduction3}\\ 
&\oplus&
\overbrace{
\begin{ytableau}
 \phantom{.} & \phantom{.} & \phantom{.}\\
 a&a
 \end{ytableau}
 \raisebox{.5em}{
$\cdots$\!
 \begin{ytableau}
 a  &  a
\end{ytableau}
}\!\!
}^{n+\bar n-2}\
\oplus  \cdots  \oplus
\overbrace{
\begin{ytableau}
 \phantom{.} & \phantom{.} \\
 a&a
 \end{ytableau}
 \raisebox{0em}{
$\cdots$\!
 \begin{ytableau}
   \phantom{.}\\
   a  
\end{ytableau}
}
 \raisebox{0.5em}{
\!\!$\cdots$\!
 \begin{ytableau}
   \phantom{.}
\end{ytableau}
}
\!\!
}^{\bar n}\ .\nonumber
\\
&&\nonumber
\end{eqnarray}
This proves our statement.

There is yet another simplification that emerges from Eq.~(\ref{reduction3}). Note that all the irreps appear only once in the reduction. That is, fixing the indices $n$, $\sigma$, and $\lb$ uniquely defines a one-dimensional subspace. Thus, the projectors $\Omega^{n,\sigma}_{\lb}$ are rank one. 

We conclude by giving explicit expressions for the dimensions of the irreps of $ S_N$ and ${\rm SU}(d)$, in Eqs.~\eqref{mult_nu_general} and~\eqref{mult_s_general}, for partitions of the form $\l=(\l_1,\l_2)$.  These expressions are  used to derive Eq.~\eqref{app:ps}, and read
\begin{eqnarray}
\nu_\l &=& \frac{N!(\l_1-\l_2+1)}{(\l_1+1)!\l_2! }\,, \label{mult_nu} \\ \nonumber\\
s_\l &=& \frac{\l_1-\l_2+1}{\l_1+1} \binom{\l_1+d-1}{d-1} \binom{\l_2+d-2}{d-2} \,. \label{mult_s}
\end{eqnarray}
One can check that Eqs.~(\ref{mult_nu}) and~(\ref{mult_s}) are consistent with Eq.~(\ref{reduction3}) by showing that the sum of the dimensions of the irreps on the right hand side agrees with the product of the two on the left hand side. Namely, by checking that
\begin{align}
s_{(\bar n,0)}s_{(n,0)}&=\sum_{i=0}^n s_{(n+\bar n-i,i)}\,,\label{check1}\\
\nu^{S_{\bar n}}_{(\bar n,0)}\nu^{S_n}_{(n,0)}\binom{n+\bar n}{n}&=\sum_{i=0}^n \nu_{(n+\bar n-i,i)}\,,\label{check2}
\end{align}
where the superscript remind us that the dimensions on the left hand side refer to irreps of  $S_{\bar n}$, $S_{n}$. One obviously obtains $\nu^{S_{\bar n}}_{(\bar n,0)}=\nu^{S_n}_{(n,0)}=1$, since these are the trivial representations of either group. The binomial in~Eq.~(\ref{check2}) arises from the definition of outer product representation in~Eq.~(\ref{reduction2}), whereby the action of $S_{n+\bar n}$ is defined on basis vectors of the form $\bar v_{i_1i_2\dots i_{\bar n}}\otimes v_{i_{\bar n+1}i_{\bar n+2}\dots i_{\bar n+n}}$, with $\bar v_{i_1i_2\dots i_{\bar n}}\in H_{\lb}^{S_{\bar n}}$, $v_{i_1i_2\dots i_{n}}\in H_{\{\l'\}}^{S_{n}}$. There are, naturally, $\binom{\bar n+n}{n}$ ways of allocating $\bar n+n$ indices in this expression.

\section{Asymptotics of $P_{\rm s}$}\label{app:asymptotics}

We next wish to address the asymptotic behavior of the success probability as the length~$N$ of the data string becomes large. Various behaviors  will be derived depending on how the local dimension $d$ scales with $N$.

In the large $N$ limit it suffices to consider even values of~$N$, which slightly simplifies the derivation of the asymptotic expressions.
The success probability in Eq.~(\ref{app:ps}) for $N=2m$, $m\in{\mathbb N}$, can be written as (just define a new index as  $j=m-i$)
\begin{equation}
\kern-.3em P_{\rm s}\!=\!{d\!-\!1\over 2^{2m-1}}\sum_{j=0}^m {(2j+1)^2\over (m\!+\!1\!+j)^2(m\!+\!d\!-\!1\!-\!j)}\!\begin{pmatrix}\!2m\\ m\!+\!j\!\end{pmatrix}\,.
\label{P asymp}
\end{equation}
For large $m$, we write $j=m x$ and use
\begin{equation}
{1\over 2^{2m-1}}\begin{pmatrix}2m\\ m+j\end{pmatrix}\sim {2\,{\rm e}^{-m x^2}\over\sqrt{m\pi}}\,.
\label{BinomialGauss}
\end{equation}
We start by assuming that $d$ scales more slowly than $N$, e.g., $d\sim N^{\gamma}$, with $0\le\gamma<1$. In this situation, we can neglect~$d$ in the denominator of Eq.~(\ref{P asymp}).
Neglecting also other subleading terms in inverse powers of $m$ and using the Euler-Maclaurin formula, we have
\begin{equation}
P_{\rm s}\sim (d-1) \int_{0}^\infty  dx\;  {4x^2\over (1+x)^2(1-x)} \;{2\,{\rm e}^{-m x^2}\over\sqrt{m\pi}} \,,
\end{equation}
which we can further approximate by substituting $0$ for~$x$ in the denominator, as the Gaussian factor peaks at $x=0$ as $m$ becomes larger, so
\begin{align}
P_{\rm s}&\sim4(d-1)\! \int_{0}^\infty  \!dx\;  x \;{2x\,{\rm e}^{-m x^2}\over\sqrt{m\pi}}\nonumber\\
&=-4(d-1) \int_{0}^\infty  dx\;  x \;{d\over dx} {{\rm e}^{-m x^2}\over m \sqrt{m\pi}}\,.
\end{align}
We integrate by parts to obtain
\begin{equation}
P_{\rm s}\sim {2(d-1)\over m}\int_{0}^\infty  dx\;  {2\,{\rm e}^{-m x^2}\over \sqrt{m\pi}}={2(d-1)\over m^2} \,.
\end{equation}
Hence, provided that $d$ scales more slowly than $N$, the probability of success vanishes asymptotically as~$N^{-2}$. More precisely, as
\begin{equation}
P_{\rm s}\sim {8(d-1)\over N^2} \,.
\label{gamma < 1}
\end{equation}

Let us next assume that $d$ scales faster than $N$, e.g., as~$d\sim N^{\gamma}$, with $\gamma>1$. In this case, $d$ is the leading contribution in the second factor in the denominator of Eq.~(\ref{P asymp}). Accordingly, we have
\begin{align}
P_{\rm s}&\sim  (d-1)m\int_0^\infty dx {4x^2\over (1+x)^2 d}{2\,{\rm e}^{-m x^2}\over\sqrt{m\pi}}\nonumber\\
&\sim 4m \int_0^\infty dx\, x {2 x\,{\rm e}^{-m x^2}\over\sqrt{m\pi}}={2\over m} \,,
\end{align}
and the asymptotic expression becomes
\begin{equation}
P_{\rm s}\sim {4\over N} \,,
\label{gamma>1}
\end{equation}
independently of $d$.

Finally, let us assume that $d$ scales exactly as $N$ and write $d=s N$, $s>0$. The success probability can be cast as
\begin{equation}
P_{\rm s}\!\sim \!(d\!-\!1)\! \int_{0}^\infty  \! dx\;  {4x^2\over (1+x)^2(1+2s-x)} \;{2\,{\rm e}^{-m x^2}\over\sqrt{m\pi}} \,.
\end{equation}
Proceeding as above, we obtain
\begin{equation}
P_{\rm s}\sim {2(d-1)\over (2s+1)m^2} \,.
\end{equation}
Thus,
\begin{equation}
P_{\rm s}\sim {8s\over (2s+1)N} \,.
\label{gamma=1}
\end{equation}
The three expressions, Eq.~(\ref{gamma < 1}), Eq.~(\ref{gamma>1}) and Eq.~(\ref{gamma=1}), can be combined into a single one as
\begin{equation}
P_{\rm s}
\sim {8(d-1)\over\left(\displaystyle 2d+N\right) N}\,.
\end{equation}

\section{Optimal POVM for general cost functions}\label{app:generalcosts}

This Appendix deals with the optimization of quantum clustering assuming other cost functions. We introduce a sufficient condition under which the type of POVM we used to maximize the success probability (Section~\ref{app:optimality}) is also optimal for a given generic cost function. We conjecture that the condition holds under reasonable assumptions. We discuss numerical results for the cases of Hamming distance, trace distance, and infidelity.

Recall that Eq.~\eqref{app:povmelements} together with Eq.~\eqref{povm} define the optimal POVM for a generic cost function that preserves covariance under $S_N$. 
However, this form is implicit and thus not very practical. 
Particularizing to the success probability, we managed to specify the function~$n(\l)=\l_2$ [cf. Eq.~\eqref{nl}] and the operators
$\Xi_\lb^n = \Omega_\lb^{n} \delta_{n,\l_2}$. In summary, the POVM was specified solely in terms of the effective states $\rho_{n,\sigma}$ (hypotheses). 

Here we conjecture that the choice \mbox{$\Xi_\lb^n=\Omega_\lb^{n} \delta_{n,n(\l)}$} is still optimal for a large class of cost functions~$f(\x,\x')$, albeit with varying guessing rules $n(\l)$. If this holds, given~$f(\x,\x')$, one only has to compute $n(\l)={\rm argmin}_n \vartheta^n_{\l,1}$ to obtain the optimal POVM. The minimum average cost can then be computed via Eq.~\eqref{opt_av_cost}. 
We now formulate this conjecture precisely as a testable mathematical condition.

For any cost function (distance) such that $f(\x,\x')\ge0$ and $f(\x,\x') = 0$ iff $\x=\x'$,
we can always find some constant $t>0$ such that 
\begin{equation}
t\,f(\x,\x')\geq \bar{\delta}_{\x,\x'}\equiv 1-\delta_{\x,\x'},\quad \forall\x,\x'.
\label{minimal cost}
\end{equation}
We can then rescale the cost function $f\mapsto t^{-1}f$ and assume with no loss of generality that $f(\x,\x')\ge\bar\delta_{\x,\x'}$.
We have 
\begin{equation}
W_\x 
= \bar W_\x + \Delta_\x ,
\end{equation}
where we have used the definition of $W_\x$ after Eq.~(\ref{Holevo1}) and similarly defined $\bar W_\x$ for the minimal cost~$\bar\delta_{\x,\x'}$. As in Section~\ref{app:optimality}, it suffices to consider $\x=(n,e)$. Then,
\begin{equation}
\Delta_\x = \sum_{\x'} \eta_{\x'}[f(\x,\x')-\bar{\delta}_{\x,\x'}] \rho_\x' \geq 0. 
\end{equation}
Using the same notation as in Eq.~(\ref{W_block}), this is equivalent~to 
\begin{equation}
\omega_\lb^n - \bar\omega_\lb^n \geq 0 \,.
\end{equation}
We now recall the meaning of Eqs.~\eqref{Holevo1b} and \eqref{Holevo2b}: the operators $\Xi_\lb^n$ must be projectors onto the eigenspace of minimal eigenvalue of $\omega_\lb^n$. Then, according to Eq.~\eqref{povm}, 
the choice $\Xi_\lb^n=\Omega_\lb^n \delta_{n,n(\l)}$
is also optimal for arbitrary cost functions if it holds that
\begin{equation}\label{conjecture}
{\rm supp}\left(\Omega_\lb^n\right) = V_1\left(\bar\omega_\lb^n\right) \overset{?}{\subset} V_1\left(\omega_\lb^n\right),
\end{equation}
where $V_1(X)$ is the eigenspace of minimal eigenvalue of~$X$, and the equality follows from Eq.~\eqref{Wme}. 

Our conjecture is that Eq.~\eqref{conjecture} holds true for the class of  ``reasonable'' cost functions considered in this paper, namely, for those that are nonnegative, covariant and satisfy the distance property stated before Eq.~(\ref{minimal cost}).  
We checked its validity for problems of size up to $N=8$, local dimension $d=2$, and uniform prior probabilities for the following cost functions:
Hamming distance $h(\x,\x') = \min\{|\x-\x'|,|\x-{\bar \x}'|\}$ ($x_i=0,1$), trace distance $T(\x,\x')=\norm{\rho_\x-\rho_{\x'}}_1$, and infidelity $I(\x,\x')=1-\tr^2\big[(\sqrt{\rho_\x} \rho_{\x'} \sqrt{\rho_\x})^{1/2}\big]$.

\begin{figure}[htbp]
\centering
	\includegraphics[scale=.47]{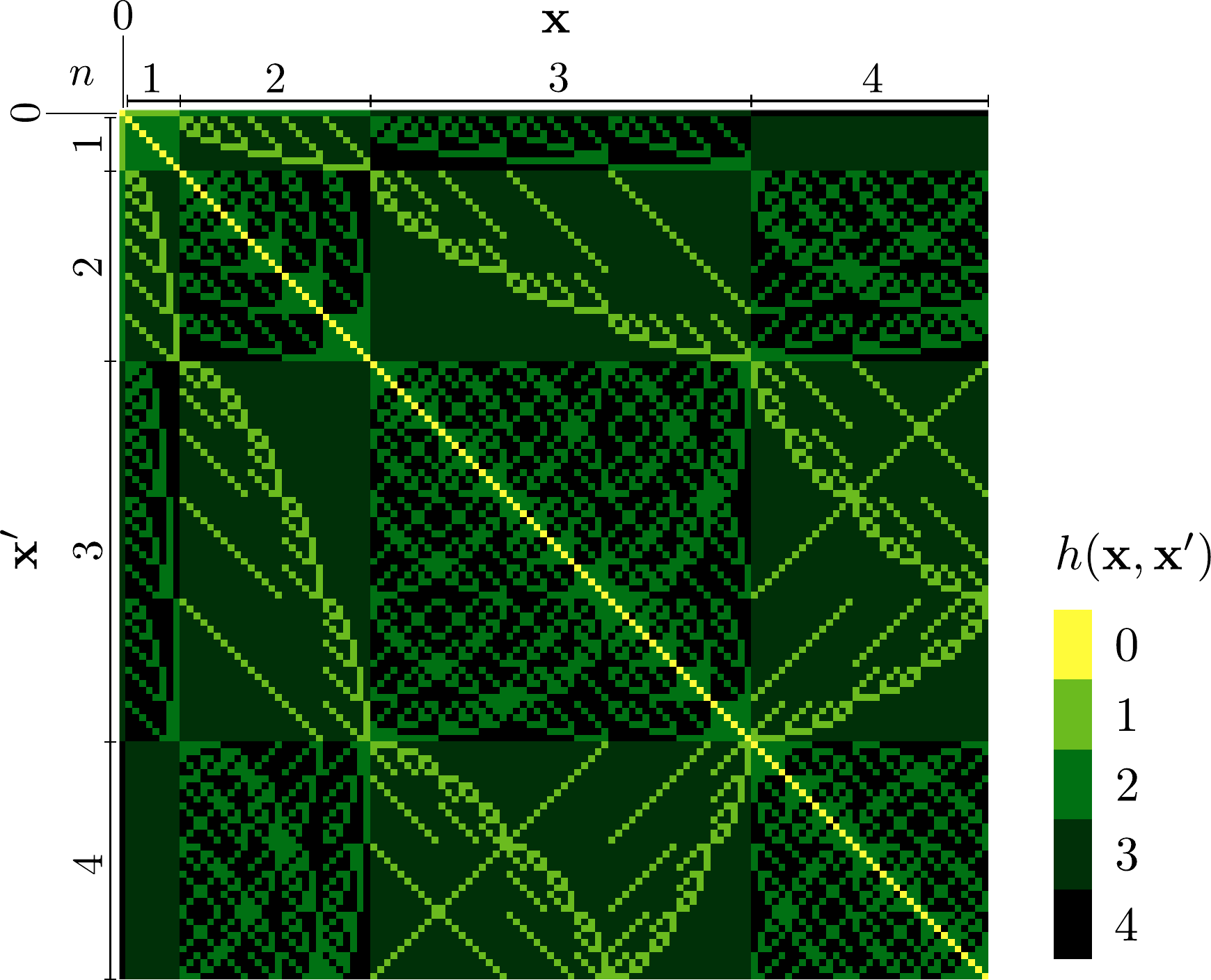}
	\caption{Heat map of the Hamming distances $h(\x,\x')$ between clusterings for $N=8$. The clusterings are grouped by size of the smallest cluster $n=0,1,2,3,4$. Each group contains all nontrivial permutations $\sigma$ for a given $n$. A brighter color means a smaller Hamming distance.}\label{fig:avocado}
\end{figure}

The above examples 
induce a much richer structure in the problem at hand. To illustrate this added complexity, in Fig.~\ref{fig:avocado} we show a heat map of the Hamming distances 
$h(\x,\x')$ between all pairs of clusterings for $N=8$. The figure shows that the largest values of $h(\x,\x')$ can occur for two clusterings with equal cluster size $n$, and that~$h(\x,\x')$ is extremely dependent on the pair of permutations $\sigma,\sigma'$. As a result, the guessing rule $n(\lambda)$ is completely different from the one that maximizes the probability of success~$P_{\rm s}$. 
In particular, irreps~$\lambda$ are no longer in one-to-one correspondence with optimal guesses for~$n$. In Table~\ref{tab:nl} we show values of~$n(\l)$ for our four cost functions and $N=4,\ldots,8$. In contrast to the case of the success probability (the cost function~$\bar\delta_{\x,\x'}$), we note that in some cases it is actually optimal to map several irreps to the same guess, while never guessing certain cluster sizes.

\begin{table*}[htbp]
\begin{tabular}{c|c|c|c|c|c|c|c|c|c|c|c|c|c|c|c|c|c|c|c|}
\cline{2-20}
                                     & \multicolumn{3}{c|}{$N=4$} & \multicolumn{3}{c|}{$N=5$} & \multicolumn{4}{c|}{$N=6$}    & \multicolumn{4}{c|}{$N=7$}    & \multicolumn{5}{c|}{$N=8$}            \\ \hline
\multicolumn{1}{|c|}{$\l$}           & (4,0)   & (3,1)   & (2,2)  & (5,0)   & (4,1)   & (3,2)  & (6,0) & (5,1) & (4,2) & (3,3) & (7,0) & (6,1) & (5,2) & (4,3) & (8,0) & (7,1) & (6,2) & (5,3) & (4,4) \\ \hline
\multicolumn{1}{|c|}{$\bar{\delta}$} & 0       & 1       & 2      & 0       & 1       & 2      & 0     & 1     & 2     & 3     & 0     & 1     & 2     & 3     & 0     & 1     & 2     & 3     & 4     \\ \hline
\multicolumn{1}{|c|}{$h$}            & 0       & 2       & 2      & 0       & 2       & 2      & 0     & 3     & 2,3   & 3     & 0     & 3     & 3     & 3     & 0     & 4     & 4     & 4     & 4     \\ \hline
\multicolumn{1}{|c|}{$T$}            & 1       & 1       & 2      & 1       & 1       & 2      & 1     & 1     & 2     & 3     & 1     & 2     & 2     & 3     & 1     & 2     & 3     & 3     & 4     \\ \hline
\multicolumn{1}{|c|}{$I$}            & 0       & 1       & 2      & 0       & 1       & 2      & 0     & 1     & 2     & 3     & 0     & 1     & 3     & 3     & 0     & 1     & 3     & 4     & 4     \\ \hline
\end{tabular}
\caption{Values of $n(\l)$, i.e., of the optimal guess for the size of the smallest cluster, where $\l=(\l_1,\l_2)$ are the relevant irreps, for data sizes $N=4,5,6,7,8$, and cost functions $\bar{\delta}(\x,\x')$ (corresponding to the success probability), Hamming distance~$h(\x,\x')$, trace distance $T(\x,\x')$, and infidelity $I(\x,\x')$.}\label{tab:nl}
\end{table*}

Performing the Schur transform is computationally inefficient on a classical computer\footnote{In contrast, as was mentioned in the main text, there exist efficient quantum circuits able to implement the Schur transform in a quantum computer. A circuit based on the Clebsch-Gordan transform achieves polynomial time in $N$ and $d$~\cite{Harrow2005}. Recently, an alternative method based on the representation theory of the symmetric group was shown to reduce the dimension scaling to ${\rm poly}(\log d)$~\cite{Krovi2018}.}, which sets a limit on the size of the data one can test---in our case it is~$N=8$. However, it is worth mentioning that this difficulty might actually be overcome. The fundamental objects needed for testing Eq.~\eqref{conjecture} are the operators $\Omega_\lb^n$. Their computation would, in principle, not require the full Schur transform, as they can be expressed in terms of generalized Racah coefficients, which give a direct relation between Schur bases arising from different coupling schemes of the tensor product space. It is indeed possible to calculate generalized Racah coefficients directly without going through a Clebsch-Gordan transform~\cite{Gliske2005}, and should this method be implemented, clustering problems of larger sizes might be tested. However, an extensive numerical analysis was not the aim of this paper.

\section{Prior distributions}\label{app:prior}

In the interest of making the paper self-contained, in this appendix we include the derivation of some results about the prior distributions used in the paper.

Let ${\mathsf S}_d=\{p_s\ge 0| \sum_{s=1}^d p_s=1\}$ denote the standard $(d-\!1)$-dimensional (probability) simplex. Every categorical distribution (CD) $P=\{p_s\}_{s=1}^d$ is a point in ${\mathsf S}_d$.
The flat distribution of CDs is the volume element divided by the volume of~${\mathsf S}_d$, the latter denoted by $V_d$. Choosing coordinates $p_1,\dots,p_{d-1}$, the flat distribution is $\prod_{s=1}^{d-1} dp_s/V_d\equiv dP$. 

Let us compute the moments of the flat distribution; as a byproduct, we will obtain $V_d$. We have
\begin{align}
V_d\!\!\int_{{\mathsf S}_d}\!\!\! dP \prod_{s=1}^d \!p_s^{n_s}
\!&=\!\!\int_0^1 \!\!dp_1\!\!\int_0^{1-p_1}\!\!\!dp_2\cdots\!\!\int_0^{1-\!\mbox{\tiny $\displaystyle\sum_{s=1}^{d-2}\!\!p_s$}}\!\!\!dp_{d-1}
\!\!\prod_{s=1}^d \!p_s^{n_s}
\nonumber\\
&=
\frac{\prod_{s=1}^d n_s!}{\left(d-1+\sum_{s=1}^d n_s\right)!}
\label{int simplex}
\end{align}
[the calculation becomes straightforward by iterating the change of variables $p_r\mapsto x$, where $p_r=(1-\sum_{s=1}^{r-1}p_s)x$, $r=d-2,d-3,\dots,2,1$]. In particular, setting $n_s=0$ for all~$s$ in Eq.~(\ref{int simplex}),  we obtain $V_d=1/(d-1)!$. Then
\begin{equation}
\int_{{\mathsf S}_d}\!\! dP \prod_{s=1}^d p_s^{n_s}=\frac{(d-1)!\prod_{s=1}^d n_s!}{\left(d-1+N\right)!},
\label{moments1}
\end{equation}
where $N=\sum_{s=1}^d n_s$.

Next, we provide a simple proof that 
any fixed von Neumann measurement on a uniform distribution of pure states in $(d,{\mathbb C})$ gives rise to CDs whose probability distribution is flat. As a result, the classical and semiclassical strategies discussed in the main text have the same success probability.

Take $|\phi\rangle\in (d,{\mathbb C})$ and let $\{|s\rangle\}_{s=1}^d$ be an orthonormal basis of $(d,{\mathbb C})$. By performing the corresponding von Neumann measurement, the probability of an outcome~$s$ is $p_s=|\langle s|\phi\rangle|^2$. Thus, any distribution of pure states induces a distribution of CDs $\{p_s=|\langle s|\phi\rangle|^2\}_{s=1}^d$ on ${\mathsf S}_d$. Let us compute the moments of the induced distribution, namely,
\begin{align}
\int\!\! d\phi \prod_{s=1}^d p_s^{n_s}\!&=
\!\!\int\!\! d\phi \,\tr\!\!\left[\bigotimes_{s=1}^d\left(|s\rangle\langle s|\right)^{\otimes n_s}\!\left(|\phi\rangle\langle\phi|\right)^{\otimes N}\!\right]\nonumber\\
&=
\frac{1}{D^{\rm sym}_N} 
\tr\!\!\left[\bigotimes_{s=1}^d\left(|s\rangle\langle s|\right)^{\otimes n_s}\!\openone^{\rm sym}_N\!\right],
\label{CalcMom}
\end{align}
where we recall that $D^{\rm sym}_N$ ($\openone^{\rm sym}_N$) is the dimension of (projector on) the symmetric subspace of~$(d,{\mathbb C})^{\otimes N}$ and we have used Schur lemma. A basis of the symmetric subspace is
\begin{equation}
|v_{\bf n}\rangle=\sqrt{{\prod_{s=1}^d n_s!\over N!}}\sum_{\sigma\in S_N}\!\!U_\sigma \bigotimes_{s=1}^d|s\rangle^{\otimes n_s},
\end{equation}
where ${\bf n}=(n_1,n_2,\dots, n_d)$. Note that there are $\binom{N+d-1}{d-1}$ different strings $\bf n$ (weak compositions of $N$ in $d$ parts), which agrees with $D^{\rm sym}_{N}=s_{(N,0)}$ [recall Ed.~(\ref{mult_s})], as it should be. Since $\openone^{\rm sym}_N=\sum_{\bf n}|v_{\bf n}\rangle\langle v_{\bf n}|$, we can easily compute the trace in Eq.~(\ref{CalcMom}) to obtain
\begin{equation}
\int\!\! d\phi \prod_{s=1}^d p_s^{n_s}\!=\!\frac{\prod_{s=1}^d n_s!}{N! D^{\rm sym}_N}=\frac{(d-1)!\prod_{s=1}^d n_s!}{(N+d-1)!}.
\label{moments2}
\end{equation}
This equation agrees with Eq.~(\ref{moments1}). This means that all the moments of the distribution induced from the uniform distribution of pure states coincide with the moments of a flat distribution of CDs on~${\mathsf S}_d$. Since the moments uniquely determine the distributions with compact support~\cite{Akhiezer1965} (and ${\mathsf S}_d$ is compact) we conclude that they are identical.

As a byproduct, we can compute the marginal distribution $\mu(c^2)$, where $c$ is the overlap of $|\phi\rangle$ with a fixed state $|\psi\rangle$. 
Since we can always find a basis such that $|\psi\rangle$ is its first element, we have $c=|\langle 1|\phi\rangle|$. Because of the results above, the marginal distribution is given by
\begin{align}
\mu(c^2)&=
\!\!\int_0^{1-p_1}\!\!dp_2\cdots\!\!\int_0^{1-\!\mbox{\tiny $\displaystyle\sum_{s=1}^{d-2}\!\!p_s$}}\!dp_{d-1}
\Bigg|_{p_1=c^2}
\nonumber\\
&=(d\!-\!1)(1\!-c^2)^{d-2},
\label{marginal}
\end{align}
in agreement with Ref.~\cite{Alonso2016}.

\section{Optimal clustering protocol for unknown classical states}\label{app:classic}

In this appendix we provide details on the derivation of the optimal protocol for a classical clustering problem, analogue to the quantum problem discussed in the main text.  The results here also apply to quantum systems when the measurement performed on each of them is restricted to be local, projective, $d$-dimensional, and fixed. We call this type of protocols semiclassical.

Here, we envision a device that takes input strings of $N$ data points $\r=(s_{1}s_2\cdots s_{N})$, with the promise that each~$s_{i}$ is a symbol out of an alphabet of $d$ symbols, say the set $\{1,2,\dots,d\}$, and has been drawn from either roulette $P$, or from roulette $Q$, with corresponding categorical probability distributions $P=\{p_{s}\}_{s=1}^{d}$ and  $Q=\{q_{s}\}_{s=1}^{d}$. 
To simplify the notation, we use the same symbols for the roulettes and their corresponding probability distributions, and  for the stochastic variables and their possible outcomes.  Also, the range of values of  the index~$s$ will always be understood to be $\{1,2,\dots, d\}$, unless specified otherwise.
The device's task is to group the data points in two clusters so that all points in either cluster have a common underlying probability distribution (either $P$ or $Q$).  We wish the machine to be universal, meaning that it shall operate without knowledge on the distributions~$P$ and~$Q$. Accordingly, we will choose as figure of merit the probability of correctly classifying \emph{all} data points, averaged over every possible  sequence of roulettes $\x=(x_1x_2\cdots x_N)$, $x_i\in\{P,Q\}$, and over every possible distribution $P$ and $Q$. The latter are assumed to be uniformly distributed over the common probability simplex ${\mathsf S}_d$ on which they are defined.
Formally, this success probability is
\begin{eqnarray}
P_{\rm s}^{\rm cl}&=&\int_{{\mathsf S}_d}\!\! dP dQ \sum_{{\x},{\r}} {\rm Pr}\left(\hat\x\in\{\x,\bar\x\},\r,\x;P,Q\right)\nonumber\\
 &=& 2 \int_{{\mathsf S}_d}\!\!  dP dQ \sum_{\x,\r} \delta_{\hat\x,\x }{\rm Pr}\left(\r,\x  ;P,Q \right),
\end{eqnarray}
where $\hat\x$ is the guess of $\x$ emitted by the machine, which by the universality requirement, can {\em only} depend on the data string $\r$.   The sums are carried out over all $2^{N}$ possible strings $\r$ and sequences of roulettes $\x$.
The factor of two in the second equality takes into account that~$P$ and~$Q$ are unknown, hence identifying the complementary string $\bar\x$ leads to the same clustering.
By emitting~$\hat\x$, the device suggests a classification of the $N$ data points~$s_i$ in two clusters.
In the above equation we have used the notation of Appendix~\ref{app:prior}
for the integral over the probability simplex.

An expression for the optimal success probability can be obtained from the trivial upper-bound
\begin{eqnarray}
P_{\rm s}^{\rm cl}&=& 
2\sum_{\r} \int dP dQ \;{\rm Pr}\left(\r,\hat\x  ;P,Q \right)\nonumber\\
&\leq& 2 \sum_{\r} \max_{\x} \int dP dQ \; {\rm Pr}\left(\r,\x  ;P,Q \right) \nonumber\\
&=& 2\sum_{\r} \max_{\x} 
 \; {\rm Pr}\left(\r,\x\right) ,
 \label{eq:Psmax}
\end{eqnarray}
where ${\rm Pr}\left(\r,\x\right)$ is the joint marginal distribution of $\r$ and~$\x$. This bound is attained by the guessing rule
\begin{equation}
 \hat\x=\underset{\x}{\operatorname{argmax}} \;{\rm Pr}\left(\r,\x\right) .
\end{equation}

For two specific distributions $P$ and $Q$, the probability that a given roulette sequence $\x$ gives rise to a particular data string $\r$ is ${\rm Pr}(\r|\x;P,Q)=\prod_{s}p_s^{n_s}q_{s}^{m_{s}}$ where $n_{s}$ ($m_{s}$) is the number of occurrences of symbol~$s$ in~$\r$ [i.e., how many $s_i\in\r$ satisfy $s_i=s$] arising from roulettes of  type $P$ ($Q$). For later convenience, we define \mbox{$M_{s}=n_{s}+m_{s}$}, which gives the total number of such occurrences. 
Note that $\{M_s\}$ is independent of $\x$, whereas~$\{n_s\}$ and $\{m_s\}$ are not.
Performing the integral over $P$ and~$Q$ we have
\begin{eqnarray}
 {\rm Pr}(\r,\x) &=& \frac{{\rm Pr}(\r|\x)}{2^{N}} \nonumber\\
 &=&\frac{1}{2^{N}}\int dP dQ\; {\rm Pr}(\r|\x;P,Q)\nonumber\\
&=& \frac{2^{-N} d_{\flat}!^{2} \prod_{s}n_{s}!m_{s}!} {(d_{\flat}+\sum_{s}m_{s})!(d_{\flat}+\sum_{s}n_{s})!} \,,
\label{eq:pxbr}
\end{eqnarray}
where we have used Eq.~(\ref{moments1}) and in the first equality we have assumed that the two types of roulette $P$ and~$Q$ are equally probable, hence each possible sequence $\x$ occurs with equal prior probability equal to $2^{-N}$. We  have also introduced the notation $d_{\flat}\equiv d-1$ to shorten the expressions throughout this appendix.  Note that all the dependence on $\x$ is through the occurrence numbers $m_s$ and $n_s$.

According to \eqref{eq:Psmax}, for each string $\r$ we need to maximize the joint probability ${\rm Pr}(\r,\x)$ in \eqref{eq:pxbr}
over all possible sequences of roulettes $\x$. We first note that, given a total of $M_s$ occurrences of a symbol $s$ in $\r$, ${\rm Pr}(\r,\x)$ is maximized by a sequence $\x$ whereby all these occurrences come from  the same type of roulette. In other words, by a sequence $\x$ such that either $m_s=M_s$ and~$n_s=0$ or else  $m_s=0$ and~$n_s=M_s$.  

In order to prove the above claim, we single out a particular symbol $r$ that occurs a total number of times $\mu=M_r$ in~$\r$. We focus on the dependence of ${\rm Pr}(\r,\x)$ 
on the occurrence number  $t=m_r$ (so, $n_r=\mu-t$) by writing
 \begin{eqnarray}
{\rm Pr}(\r,\x)&=&
 \frac{a \,(\mu-t)! t!}{(b+t )!(c-t)!} \equiv f(t) ,
 \end{eqnarray}
where the coefficients $a$, $b$, and $c$ are defined as
 \begin{eqnarray}
 a&=&\frac{d_{\flat}!^{2}}{2^{N}}\prod_{s\neq r} n_{s}!m_{s}!\,,\\
 b&=&d_{\flat}+\sum_{s\neq r} m_s\,,\\
 c&=&d_{\flat}+\sum_{s} n_s+m_r=d_{\flat}+N-\sum_{s\neq r} m_s\,,
 \label{the C}
 \end{eqnarray}
and are independent of $t$. The function $f(t)$ can be extended to $t\in{\mathbb R}$ using the Euler gamma function and the relation $\Gamma(t+1)=t!$.
This enables us to compute the second derivative of $f(t)$ and show that it is a convex function of $t$ in the interval $[0,\mu]$. Indeed, 
 \begin{align}\label{harmonic}
\kern-1em{f''(t)\over f(t)}&=\!\left[H_1(c\!-\!t)\!-\!H_1(\mu\!-\!t)   
\!-\!H_1(b\!+\!t)\!+\!  H_1(t)  \right]^2\nonumber\\
&+\! \phantom{\left[\right.}H_2(c\!-\!t)\!-\!H_2(\mu\!-\!t)
\!+\! H_2(b\!+\!t)\!-\!H_2(t)          \nonumber\\[.5em]
&\geq 0\,, 
 \end{align}
where $H_n(t)$ are the generalized harmonic numbers. For positive integer values of $t$ they are $H_n(t)=\sum_{j=1}^{t} j^{-n}$. The relation $ H_n(t)=\zeta(n)-\sum_{j=1}^\infty (t+j)^{-n}$, where $\zeta(n)=\sum_{j=1}^\infty j^{-n}$ is the Riemann zeta function, allows to extend the domain of $H_n(t)$ to real (and complex) values of $t$. 

The positivity of $f''(t)$  follows from the positivity of both~$f(t)$ and the two differences of harmonic numbers in the second line of Eq.~(\ref{harmonic}). Note that $H_2(x)$ is an increasing function of $x$. Since, obviously, $b+t>t$, and $c-t>\sum_s n_s=\sum_s(M_s-m_s)\ge \mu-t$ [as follows from the definition of $c$ in~Eq.~(\ref{the C})], we see that the two differences are positive. 

The convexity of $f(t)$ for $t\in[0,\mu]$ implies that the maximum of $f(t)$ is either at $t=0$ or $t=\mu$. This holds for every value of  $M_r $ and every symbol $r$ in the data string, so our claim holds.  In summary, the optimal guessing rule must assign  the same type of roulette to all the $M_s$ occurrences of a symbol $s$, i.e., it must group all data points that show the same symbol in the same cluster. This is in full agreement with our own intuition. 

The description of the optimal protocol that runs on our device is not yet complete. We need to specify how to reduce the current number of clusters down to two, since at this point we may (and typically will) have up to $d$ clusters; as many as different symbols. The reduction, or merging of the $d$ clusters can only be based on their relative sizes, as nothing is known about the underlying probability distributions. 
This is quite clear: Let ${\mathsf P}$ be the subset of symbols (e.g., the subset of $\{1,2,\dots,d\}$) for which $n_s=M_s$, and let ${\mathsf Q}$ be its complement, i.e.,~${\mathsf Q}$ contains the symbols for which $m_s=M_s$, and ${\mathsf P}=\bar{\mathsf Q}$. The claim we just proved tells us that in order to find the maximum of ${\rm Pr}(\r,\x)$ it is enough to consider sequences of roulettes $\x$ that comply with the above conditions on the occurrence numbers.\footnote{For example, suppose $d=3$ and $N=12$. Assuming that $\r=(112321223112)$ is the string of data, the sequence of roulettes $\x$ in the table
$$
\begin{tabular}{c | c c c c c c c c c c c c c} 
  $i$ &1&2&3&4&5&6&7&8&9&10&11&12 \\ [0.5ex] 
  \hline
 $\r$ &1&1&2&3&2&1&2&2&3&1&1&2 \\ [0.5ex] 
 \hline
$\x$ &$P$&$P$&$Q$&$Q$&$Q$&$P$&$Q$&$Q$&$Q$&$P$&$P$&$Q$
  \end{tabular}
 $$
 satisfies the conditions $m_s=M_s$ or $n_s=M_s$, since
 $n_1=M_1=5$, $m_2=M_2=5$, and $m_3=M_3=2$. In this case, ${\mathsf P}=\{1\}$, and ${\mathsf Q}=\{2,3\}$. 
 The suggested  clustering is $\{(1,2,6,10,11),(3,4,5,7,8,9,12)\}$.
} For those, the joint probability~${\rm Pr}(\r,\x)$ can be written as
\begin{equation}
{\rm Pr}(\r,\x)=
 \frac{a}{ \big(d_{\flat}+\sum_{s\in{\mathsf Q}} M_s \big)!\big(d_{\flat}+\sum_{s\in{\mathsf P}} M_s\big)!}\,,
 \label{eq:maxpsx-1}
\end{equation}
where $a$ now simplifies to $ 2^{-N} d_{\flat}!^{2}\raisebox{.15em}{\small${\prod}_{s}$} M_s!$.
Thus, it just remains to find the partition $\{{\mathsf P},{\mathsf Q}\}$ that maximizes this expression. It can be also be written as
\begin{equation}
{\rm Pr}(\r,\x)=
 \frac{a}{(d_{\flat}+x )!(d_{\flat}+N-x)!}\,,
 \label{eq:maxpsx}
\end{equation}
where we have defined $x=\raisebox{.15em}{\small$\sum_{s\in {\mathsf Q}}$} M_s$. 
The maximum of this function is located at $x=N/2$, and one can easily check that it is monotonic on either side of its peak.
Note that, depending on the values of the occurrence numbers $\{M_{s}\}$, the optimal value, $x=N/2$, may not be attained. In such cases, the maximum of ${\rm Pr}(\r,\x)$  is located at $x^*=N/2\pm\Delta$,  where $\Delta$ is the bias 
\begin{equation}
\Delta=\frac{1}{2}\min_{\mathsf Q}\left|\sum_{s\in {\mathsf Q}}M_s-\sum_{s\in\bar{\mathsf Q}}M_s\right|\,.
\label{eq:bias}
\end{equation}
The subset $\mathsf Q$ that minimizes this expression determines the optimal clustering.

In summary (and not very surprisingly), the optimal guessing rule consists in first partitioning the data $\r$ in up to $d$ groups according to the symbol of the data points, and secondly, merging  those groups (without splitting them) in two clusters in such a way that their sizes are as similar as possible.
We have stumbled upon the so-called \emph{partition problem}~\cite{Korf1998}, which is known to be weakly NP-complete. In particular, a large set of distinct occurrence counts $\{M_s\}$ rapidly hinders the efficiency of known algorithms, a situation likely to occur for large $d$. It follows that the optimal clustering protocol for the classical problem cannot be implemented efficiently in all instances of the problem.

To obtain the maximum success probability $P_{\rm s}^{\rm cl}$,  Eq.~(\ref{eq:Psmax}), we need to sum the maximum joint probability, given by \eqref{eq:maxpsx} with $x=x^*$, over all
possible strings~$\r$. Those with the same set of occurrence counts~$\{M_{s}\}$ 
give the same contribution. Moreover, all the dependence on~$\{M_s\}$ is through the bias $\Delta$. Therefore, if we define $\xi_\Delta$ to be the number of sets $\{M_s\}$ that give rise to a bias $\Delta$, then the corresponding number of data strings is $\xi_\Delta N!/\raisebox{.15em}{\small${\prod}_{s}$} M_s!$. We thus can write
\begin{equation}
P_{s}^{\rm cl}=\sum_{\Delta}
 \frac{2^{1-N}\xi_\Delta d_{\flat}!^{2}N!}{\left(d_{\flat}\!+\!{N\over2}\!+\!\Delta \right)!\left(d_{\flat}\!+\!{N\over2}\!-\!\Delta\right)!}\,.
 \label{eq:Pssum}
\end{equation}

This is as far as we can go, as no explicit formula for the combinatorial factor $\xi_\Delta$ is likely to exist for general cases. However, it is possible to work out the asymptotic expression of the maximum success probability for large data sizes $N$. 
We first note that a generic term in the sum~(\ref{eq:Pssum}) can be written as the factor $2^{2d_\flat+1}\xi_\Delta d_\flat!^2 N!/(2d_\flat+N)!$ times a binomial distribution that peaks at $\Delta=0$ for large $N$. Hence, the dominant contribution in this limit is
 \begin{eqnarray}
P_{s}^{\rm cl}&\sim &\xi_{0}\frac{2^{2d_\flat+1} d_{\flat}!^{2}N!}{(2d_{\flat}+N)!}
\sim \xi_{0}\frac{2^{2d-1}(d-1)!^{2}}{N^{2d-2}} .
 \label{eq:Psfopt2}
 \end{eqnarray}

From the definition of $\xi_\Delta$, given above Eq.~(\ref{eq:Pssum}), and that of $\Delta$ in Eq.~(\ref{eq:bias}), we readily see that $\xi_{0}$ is the number of ordered partitions (i.e., the order matters) of~$N$ in~$d$ addends or parts\footnote{These ordered partitions 
are known as {\em weak compositions} of $N$ into $d$ parts in combinatorics, where {\em weak} means that some addends (or parts) can be zero; in contradistinction, the term {\em composition} is used when all the parts are strictly positive. 
}  (the occurrence counts~$M_s$) such that a subset of these addends 
is an ordered partition of~$N/2$ as well. 

\begin{figure}[htbp]
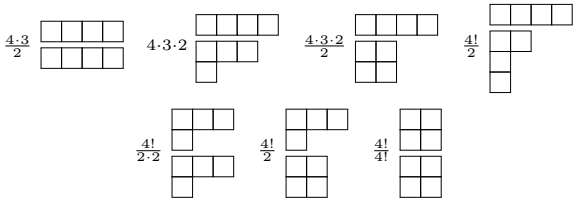

\ytableausetup{mathmode,boxsize=.8em,aligntableaux=top}
\begin{gather*}
\scriptstyle\frac{4\cdot3}{2}
\begin{array}{l}
\ydiagram{4}\\[-.3em]
 \ydiagram{4}
 \end{array}
 \phantom{+}
 \scriptstyle4\cdot3\cdot2
 \begin{array}{l}
\ydiagram{4}\\[-.3em]
 \ydiagram{3,1}
 \end{array}
 \phantom{+}
 \scriptstyle{4\cdot3\cdot2\over2}
  \begin{array}{l}
\ydiagram{4}\\[-.3em]
 \ydiagram{2,2}
 \end{array}
  \phantom{+}\scriptstyle {4!\over2}
  \begin{array}{l}
\ydiagram{4}\\[-.3em]
 \ydiagram{2,1,1}
 \end{array}\nonumber\\
 \scriptstyle\frac{4!}{2\cdot2}
 \begin{array}{l}
\ydiagram{3,1}\\
 \ydiagram{3,1}
 \end{array}
 \phantom{+} \scriptstyle\frac{4!}{2}
 \begin{array}{l}
\ydiagram{3,1}\\
 \ydiagram{2,2}
 \end{array}
  \phantom{+}
   \scriptstyle \scriptstyle {4!\over4!}
  \begin{array}{l}
\ydiagram{2,2}\\
 \ydiagram{2,2}
 \end{array}
\end{gather*}
\caption{Use of Young diagrams for computing $\xi_0$. In the example, $N=8$ and $d=4$. The fraction before each pair gives the number of different fillings and hints at how it has been computed.}
\label{fig:counting}
\end{figure}

Young diagrams come in handy to compute~$\xi_0$. First, we draw pairs of diagrams, $[\l,\l']$,  each of $N/2$ boxes and such that $\l\ge\l'$ (in lexicographical order; see Appendix~\ref{app:partitions}), and $l(\l)+l(\l')\equiv r+r'\le d$, i.e., the total number of rows should not exceed $d$. Next, we fill the boxes with symbols $s_i$ (representing possible data points) so that all the boxes in each row have the same symbol. We readily see that the number of different fillings gives us $\xi_0$.  
An example is provided in Fig.~\ref{fig:counting} for clarity.

Although this pictorial method eases the computation of $\xi_0$, it becomes unpractical even for relatively small values of $N$. However, it becomes again very useful in the asymptotic limit since the number of Young diagrams with  at least two rows of equal size become negligibly small for large $N$.\footnote{Actually,  the number of Young diagrams of a given length with unequal number of boxes in each row is equal to the number of Young diagrams of $N-r(r-1)/2$ boxes, i.e., it is equal to $P^{(r)}_{N-r(r-1)/2}$. Using the results in Appendix~\ref{app:partitions},  we immediately see that  for large $N$ one has $P^{(r)}_{N-r(r-1)/2}/P^{(r)}_N\sim 1$, which proves the statement.} The same conclusion applies to the whole pairs $[\l,\l']$, since e.g., by reshuffling rows, one could merge the two members into  a single diagram of~$N$ boxes and length $r+r'$.  Thus, we may assume that all pairs of diagrams with a given total length, have unequal number of boxes in each row, which renders the counting of different fillings trivial: there are \mbox{$d!/(d-r-r'+1)!$} ways of filling each pair of diagrams. 
Recalling that there is a one-to-one mapping between partitions and Young diagrams, we can use Eq.~(\ref{partAsym}) and write
\begin{align}
\xi_{0}&\sim\frac{1}{2}\sum_{r=1}^{d-1}\sum_{r'=1}^{d-r} P^{(r)}_{N\over2}P^{(r')}_{N\over2}\frac{d!}{(d-r-r')!}\nonumber\\
&\sim\frac{1}{2} \left(\frac{N}{2}\right)^{d-2}\sum_{r=1}^{d}\frac{r (d-r) d!}{r!^{2}(d-r)!^{2}}\nonumber\\
&\sim\frac{1}{2} \left(\frac{N}{2}\right)^{d-2} \frac{(2d-2)!}{(d-2)!(d-1)!^{2}}\,.
\end{align}
This result, together with \eqref{eq:Psfopt2}, leads us to the desired asymptotic expression for the optimal success probability:
\begin{equation}
P_{s}^{\rm cl} \sim  \left(\frac{2}{N}\right)^{d} \frac{(2d-2)!}{(d-2)!}\,.
\label{eq:PsfoptFinal}
\end{equation}

\blue{
\section{Optimal clustering protocol for known classical states}\label{app:classic_known}

In this Appendix, we give a short discussion on clustering classical states under the assumption that the underlying probability distributions are known. In particular, we discuss two low-dimensional cases, $d=2,3$, and derive the asymptotic expression of the success probability of clustering for large data string length $N$ and arbitrary data dimension~$d$.
We stick to the notation introduced in Appendix~\ref{app:classic}. 

If the underlying probability distributions are known, a given data point $s$ is optimally assigned to the probability distribution for which $s$ is most likely. The success probability is thus given by $\max\{p_s,q_s\}/2$ (recall that the data is assumed to be drawn from either $P$ or $Q$ with equal prior probability). 
The average success probability of clustering over all possible strings of length~$N$ then reads
%
%
\begin{equation}
\label{KP-2}
\kern-0.4em P^{\rm cl}_{{\rm s},PQ}\!
=\!\frac{1}{2^N}\!\!\left[\!\left(\sum_{s=1}^d\! \max\{p_s,q_s\}\!\!\right)^{\!\!N}\!\!\!\!+\!
\left(\sum_{s=1}^d \!\min\{p_s,q_s\}\!\!\right)^{\!\!N}\right]\!\!,\!\!
\end{equation}
where the term in the second line arises because assigning the wrong probability distribution to {\em all} data points in~$\r$ gives a correct clustering. 
In order to compare with our results for unknown classical states, we average the success probability over a uniform distribution of categorical probability distributions. This yields
\begin{equation}
\label{KP-3}
P^{\rm cl}_{\rm s}=\int_{{\mathsf S}_d}\!\!  dP \int_{{\mathsf S}_d}\!\!  dQ \,P^{\rm cl}_{{\rm s},PQ}\,,
\end{equation}
where the integration over the simplex ${\mathsf S}_d$, shared by $P$ and $Q$,
is defined in Appendix~\ref{app:prior}.

To perform the integral in Eq.~(\ref{KP-3}) we need to partition ${\mathsf S}_d\times{\mathsf S}_d$ in different regions according to whether \mbox{$p_s\le q_s$} or $p_s>q_s$ for the various symbols.
By symmetry, the integral can only depend on the number $r$ of symbols for which $p_s\le q_s$ (not in its particular value).
%
%
Hence, $r=1,\dots,d-1$ labels the different types of integrals that we need to compute to evaluate~$P_{\rm s}^{\rm cl}$. Notice that we have the additional symmetry $r\leftrightarrow d-r$, corresponding to exchanging $p_s$ and $q_s$ for all~$s$.
Since the value of these integrals does not depend on the specific value of~$s$, we can choose all $p_s$ with $s=1,2,\dots,r$ to satisfy $p_s>q_s$ and all $p_s$ with $s=r+1,r+2,\dots, d$ to satisfy \mbox{$p_s\le q_s$}.
%
To shorten the expressions below, we define
\begin{equation}
\label{KP-6}
{\mathfrak p}_k:=\sum_{s=1}^k
p_s\,,\quad
{\mathfrak q}_k:=\sum_{s=1}^k
q_s \,.
\end{equation}
With these definitions ${\mathfrak p}_d={\mathfrak q}_d=1$, $\sum_{s=r+1}^dq_s=1-{\mathfrak q}_{r}$, and likewise \raisebox{0ex}[0ex][0ex]{$\sum_{s=r+1}^dp_s=1-{\mathfrak p}_{r}$}.
The integrals that we need to compute are then
%
\begin{multline}
\label{KP-8}
I^d_r:=\!\! \int_{{\mathsf S}_d} \!\!\!dP\,
{1\over V_d}\!\int_{0}^{p_1}\!\!\! dq_1 \cdots\!\!
\int_{0}^{p_r}\!\!\! dq_r \\
\times \int_{p_{r+1}}^{{\mathfrak p}_{r+1}\!-{\mathfrak q}_r}\!\!\!\! dq_{r+1} \cdots\!\!
\int_{p_{d-1}}^{{\mathfrak p}_{d-1}\!-{\mathfrak q}_{d-2}}\!\!\!\!  dq_{d-1} \\
 \times\left[ (1\!+\!{\mathfrak p}_r\!-\!{\mathfrak q}_r)^N\!\!+(1\!+\!{\mathfrak q}_r\!-\!{\mathfrak p}_r)^N \right],
\end{multline} 
%
and we note that, as anticipated, $I^d_r=I^d_{d-r}$. The average probability of successful clustering then reads 
\begin{equation}
\label{KP-11}
P^{\rm cl}_{\rm s}=\frac{1}{2^N}\sum_{r=1}^{d-1} \binom{d}{r}I^d_r,
\end{equation}
where the binomial is the number of equivalent integral regions for the given $r$.

\subsubsection*{Low data dimension}

We can now discuss the lowest dimensional cases, for which explicit closed formulas for $I^d_r$ can be derived. For $d=2$ one has
\begin{equation}
\label{KP-12}
P^{\rm cl}_{\rm s}=\frac{8-2^{2-N}}{(N+2)(N+1)}.
\end{equation}
This result coincides with that of unknown probability distributions given in Eq.~\eqref{eq:Pssum} with $\xi_\Delta=1$. This is an expected result, as the optimal protocol for known and unknown probability distributions is exactly the same: assign to the same cluster all data points that show the same symbol~$s$. Therefore, knowing the probability distribution does not provide any advantage for $d=2$.

For $d>2$, however, knowledge of the distributions $P$ and $Q$ helps classifying the data points. If $d=3$, the success probability \eqref{KP-11} can be computed to be
\begin{equation}
\label{KP-13}
P^{\rm cl}_{\rm s}=6\frac{2^5 (N-2)-2^{2-N}(N^2+7N+18)}{(N+4)(N+3)(N+2)(N+1)}\,.
\end{equation}
In Table~\ref{table:K-U} we compare five values of $P^{\rm cl}_{\rm s}$ in Eq~\eqref{KP-13}, when $N=2,3,\dots,6$,  with those for unknown distributions $P$ and $Q$ given by Eq.~\eqref{eq:Pssum}. As expected, the success probability is larger if $P$ and $Q$ are known. The source of the increase is illustrated by  the string $\r=(112)$, which would be labeled as $PPQ$ (or $QQP$) if $P$ and $Q$ were unknown. However,  if they are known and, e.g.,  $p_1>q_1$ and $p_2>q_2$, the string will be more appropriately labeled as $PPP$. 
%

\begin{table}
\blue{
\begin{tabular}{|c|c|c|c|c|c|}
\hline
$N$ & 2 &3 &4 &5 & 6\\
\hline\hline
Unknown: & 7/12 & 11/30 & 0.250 & 0.176 & 0.130\\
\hline
Known: & 3/5 & 2/5 & 0.283 & 0.210 & 0.160  \\
\hline
\end{tabular}
\caption{The success probability $P^{\rm cl}_{\rm s}$ for $d=3$ and data string lengths $N=2,\dots,6$ in the cases of known and unknown distributions $P$ and $Q$. For unknown distributions, the values are computed using Eq.~\eqref{eq:Pssum}  in  Appendix~\ref{app:classic}. For known distributions, the values are given by Eq.~\eqref{KP-13}. The table shows that knowing $P$ and $Q$ increases the success probability of clustering.}
\label{table:K-U}
}
\end{table}

\subsubsection*{Arbitrary data dimension. Large $N$ limit}

For increasing $N$, however, the advantage of knowing $P$ and $Q$ becomes less significant and vanishes asymptotically.  This can be checked explicitly for $d=2,3$ by expanding Eqs.~(\ref{KP-12}) and~(\ref{KP-13}) in inverse powers of $N$. In this regime the average is dominated by distributions for which ${\mathfrak p}_r \approx 1$ and ${\mathfrak q}_r \approx 0$. Since in a typical string 
approximately half of the data will come from the distribution~$P$ and the other half from~$Q$, the optimal clustering protocol will essentially coincide with that for unknown distributions, i.e., it will collect the data points showing the same symbol in the same subcluster and afterwards merge the subclusters into two clusters of approximately the same size. We next prove that this intuition is right for all $d$.

In the proof, we will make repeated use of the trivial observation that, for asymptotically large $N$ and $0<a<b<c$, one has 
\begin{equation}\label{KP-ebc2}
\int_a^b (c-x)^N dx \sim (c-a)^{N+1} /N\,. 
\end{equation}
We also
note that the contribution to the success probability coming from the completely wrong assignment of distributions, i.e., $(1+{\mathfrak q}_r-{\mathfrak p}_r)^N$,  is exponentially vanishing, since we assumed $p_r>q_r$, and thus ${\mathfrak q}_r-{\mathfrak p}_r<0$ [this is well illustrated by the terms proportional to $2^{2-N}$ in Eqs.~(\ref{KP-12}) and~(\ref{KP-13})]. 

Because of this last observation, we can drop the second term in the integrand of $I^d_r$, Eq.~(\ref{KP-8}).
The integrals over~$q_s$, $s\le r$, of the remaining term, $(1+{\mathfrak p}_r-{\mathfrak q}_r)^N$, 
are dominated by the lower limit, $q_s=0$, as this value maximizes $1+{\mathfrak p}_r-{\mathfrak q}_r$. Using Eq.~(\ref{KP-ebc2}) we get
\begin{multline}
\label{KP-ebc1}
I^d_r\sim{(d-1)!\over N^r}\int_{{\mathsf S}_d} \!\!\!dP\\ 
\times\int_{p_{r+1}}^{{\mathfrak p}_{r+1}\!-{\mathfrak q}_r}\!\!\!\! dq_{r+1} \cdots\!\!
\int_{p_{d-1}}^{{\mathfrak p}_{d-1}\!-{\mathfrak q}_{d-2}}\!\!\!\!  dq_{d-1}  (1\!+\!{\mathfrak p}_r)^{N+r}\,,
\end{multline}
where we recalled that the volume of the simplex ${\mathsf S}_d$ is $V_d=1/(d-1)!$.
For the remaining integrals over $q_s$ 
in Eq.~(\ref{KP-ebc1}) we can take the lower limits to be \mbox{$p_s\approx 0$}, for \mbox{$s\ge r+1$}, since the integrand is maximized by \mbox{${\mathfrak p}_r\approx 1$}. Therefore, the upper limits become $1$, $1-q_{r+1}$, \dots, $1-\sum_{s=r+1}^{d-2}q_s$. We identify this upper and lower limits as those of an integral over a $(d-r-1)$-dimensional probability simplex ${\mathsf S}_{d-r}$. We can thus write
%
\begin{equation}
\label{KP-as-1}
I_r^d\sim \frac{(d-1)!}{(d-r-1)! N^r} \int_{{\mathsf S}_d} \!\!dP\,(1+{\mathfrak p}_r)^{N+r} \,.
\end{equation}

The last equation can be cast as
\begin{equation}
\label{KP-as-1}
I_r^d\sim \frac{(d-1)!}{(d-r-1)! N^r} \int_{{\mathsf S}_d} \!\!dP\left(\!2\!-\!\sum_{s=r}^{d-1} p_s\!\right)^{\!N+r}\,,
\end{equation}
where we have used again that ${\mathfrak p}_r=1-\sum_{s=r+1}^dp_s$ and noted that under the integral sign we are free to relabel the variables $p_s$.
%
According to the definition of~$\int_{\raisebox{.3ex}[0ex][0ex]{\tiny${\mathsf S}_d$}}\!dP$, we need to perform $d-r$ integrals over the variables $p_{r},p_{r+1},\cdots, p_{d-1}$, for which we can use Eq.~(\ref{KP-ebc2}). This yields  a factor $2^{N+d}/N^{d-r}$. The remaining integrals over~$p_1,p_2,\dots,p_{r-1}$  of this constant factor give an additional $1/(r-1)!$, as they effectively correspond to an integral over a $(r-1)$-dimensional simplex. 
Putting the different pieces together, the asymptotic expression of $I^d_r$ reads
\begin{equation}
\label{KP-as-2}
I_r^d\simeq \frac{2^{N+d}}{N^d}\frac{[(d-1)!]^2}{(r-1)! (d-r-1)!}\,.
\end{equation}

We are now in position to compute the asymptotic success probability. Inserting Eq.~\eqref{KP-as-2} into Eq.~\eqref{KP-11} we readily obtain
\begin{align}
\label{KP-as-3}
P^{\rm cl}_{\rm s}&\sim 
      \left(\frac{2}{N}\right)^d(d\!-\!1)!(d\!-\!1)\sum_{r=1}^{d-1} \binom{d}{r}\binom{d\!-\!2}{d\!-\!r\!-\!1} \nonumber\\
     & = \left(\frac{2}{N}\right)^d\frac{(2d-2)!}{(d-2)!}\,,
\end{align}
where we have used the well-known binomial  identity $\sum_k\binom{a}{k}\binom{b}{s-k}=\binom{a+b}{s}$ [here, $k$ ranges over all values for which the binomials make sense]. 
Eq.~\eqref{KP-as-3} coincides with the asymptotic expression in the unknown case Eq.\eqref{ps_asym_cl}, as we anticipated.
}

\bibliography{_Unsupervised}

\begin{thebibliography}{62}%
\makeatletter
\providecommand \@ifxundefined [1]{%
 \@ifx{#1\undefined}
}%
\providecommand \@ifnum [1]{%
 \ifnum #1\expandafter \@firstoftwo
 \else \expandafter \@secondoftwo
 \fi
}%
\providecommand \@ifx [1]{%
 \ifx #1\expandafter \@firstoftwo
 \else \expandafter \@secondoftwo
 \fi
}%
\providecommand \natexlab [1]{#1}%
\providecommand \enquote  [1]{``#1''}%
\providecommand \bibnamefont  [1]{#1}%
\providecommand \bibfnamefont [1]{#1}%
\providecommand \citenamefont [1]{#1}%
\providecommand \href@noop [0]{\@secondoftwo}%
\providecommand \href [0]{\begingroup \@sanitize@url \@href}%
\providecommand \@href[1]{\@@startlink{#1}\@@href}%
\providecommand \@@href[1]{\endgroup#1\@@endlink}%
\providecommand \@sanitize@url [0]{\catcode `\\12\catcode `\$12\catcode
  `\&12\catcode `\#12\catcode `\^12\catcode `\_12\catcode `\%12\relax}%
\providecommand \@@startlink[1]{}%
\providecommand \@@endlink[0]{}%
\providecommand \url  [0]{\begingroup\@sanitize@url \@url }%
\providecommand \@url [1]{\endgroup\@href {#1}{\urlprefix }}%
\providecommand \urlprefix  [0]{URL }%
\providecommand \Eprint [0]{\href }%
\providecommand \doibase [0]{http://dx.doi.org/}%
\providecommand \selectlanguage [0]{\@gobble}%
\providecommand \bibinfo  [0]{\@secondoftwo}%
\providecommand \bibfield  [0]{\@secondoftwo}%
\providecommand \translation [1]{[#1]}%
\providecommand \BibitemOpen [0]{}%
\providecommand \bibitemStop [0]{}%
\providecommand \bibitemNoStop [0]{.\EOS\space}%
\providecommand \EOS [0]{\spacefactor3000\relax}%
\providecommand \BibitemShut  [1]{\csname bibitem#1\endcsname}%
\let\auto@bib@innerbib\@empty
\bibitem [{\citenamefont {Shor}(1996)}]{Shor1998}%
  \BibitemOpen
  \bibfield  {author} {\bibinfo {author} {\bibfnamefont {Peter~W.}\
  \bibnamefont {Shor}},\ }\bibfield  {title} {\enquote {\bibinfo {title}
  {{Fault-tolerant quantum computation}},}\ }in\ \href {\doibase
  10.1109/SFCS.1996.548464} {\emph {\bibinfo {booktitle} {Proceedings of 37th
  Conference on Foundations of Computer Science}}}\ (\bibinfo  {publisher}
  {IEEE Comput. Soc. Press},\ \bibinfo {year} {1996})\ pp.\ \bibinfo {pages}
  {56--65}\BibitemShut {NoStop}%
\bibitem [{\citenamefont {Grover}(1997)}]{Grover1997}%
  \BibitemOpen
  \bibfield  {author} {\bibinfo {author} {\bibfnamefont {Lov~K.}\ \bibnamefont
  {Grover}},\ }\bibfield  {title} {\enquote {\bibinfo {title} {{Quantum
  Mechanics Helps in Searching for a Needle in a Haystack}},}\ }\href {\doibase
  10.1103/PhysRevLett.79.325} {\bibfield  {journal} {\bibinfo  {journal}
  {Physical Review Letters}\ }\textbf {\bibinfo {volume} {79}},\ \bibinfo
  {pages} {325--328} (\bibinfo {year} {1997})},\ \Eprint
  {http://arxiv.org/abs/9706033} {arXiv:9706033 [quant-ph]} \BibitemShut
  {NoStop}%
\bibitem [{\citenamefont {Finnila}\ \emph {et~al.}(1994)\citenamefont
  {Finnila}, \citenamefont {Gomez}, \citenamefont {Sebenik}, \citenamefont
  {Stenson},\ and\ \citenamefont {Doll}}]{Finnila1994}%
  \BibitemOpen
  \bibfield  {author} {\bibinfo {author} {\bibfnamefont {A.B.}\ \bibnamefont
  {Finnila}}, \bibinfo {author} {\bibfnamefont {M.A.}\ \bibnamefont {Gomez}},
  \bibinfo {author} {\bibfnamefont {C.}~\bibnamefont {Sebenik}}, \bibinfo
  {author} {\bibfnamefont {C.}~\bibnamefont {Stenson}}, \ and\ \bibinfo
  {author} {\bibfnamefont {J.D.}\ \bibnamefont {Doll}},\ }\bibfield  {title}
  {\enquote {\bibinfo {title} {{Quantum annealing: A new method for minimizing
  multidimensional functions}},}\ }\href {\doibase
  10.1016/0009-2614(94)00117-0} {\bibfield  {journal} {\bibinfo  {journal}
  {Chemical Physics Letters}\ }\textbf {\bibinfo {volume} {219}},\ \bibinfo
  {pages} {343--348} (\bibinfo {year} {1994})},\ \Eprint
  {http://arxiv.org/abs/9404003} {arXiv:9404003 [chem-ph]} \BibitemShut
  {NoStop}%
\bibitem [{\citenamefont {Kadowaki}\ and\ \citenamefont
  {Nishimori}(1998)}]{Kadowaki1998}%
  \BibitemOpen
  \bibfield  {author} {\bibinfo {author} {\bibfnamefont {Tadashi}\ \bibnamefont
  {Kadowaki}}\ and\ \bibinfo {author} {\bibfnamefont {Hidetoshi}\ \bibnamefont
  {Nishimori}},\ }\bibfield  {title} {\enquote {\bibinfo {title} {{Quantum
  annealing in the transverse Ising model}},}\ }\href {\doibase
  10.1103/PhysRevE.58.5355} {\bibfield  {journal} {\bibinfo  {journal}
  {Physical Review E}\ }\textbf {\bibinfo {volume} {58}},\ \bibinfo {pages}
  {5355--5363} (\bibinfo {year} {1998})},\ \Eprint
  {http://arxiv.org/abs/9804280} {arXiv:9804280 [cond-mat]} \BibitemShut
  {NoStop}%
\bibitem [{\citenamefont {Lloyd}(1996)}]{Lloyd1996}%
  \BibitemOpen
  \bibfield  {author} {\bibinfo {author} {\bibfnamefont {S.}~\bibnamefont
  {Lloyd}},\ }\bibfield  {title} {\enquote {\bibinfo {title} {{Universal
  Quantum Simulators}},}\ }\href {\doibase 10.1126/science.273.5278.1073}
  {\bibfield  {journal} {\bibinfo  {journal} {Science}\ }\textbf {\bibinfo
  {volume} {273}},\ \bibinfo {pages} {1073--1078} (\bibinfo {year}
  {1996})}\BibitemShut {NoStop}%
\bibitem [{\citenamefont {Kimble}(2008)}]{Kimble2008}%
  \BibitemOpen
  \bibfield  {author} {\bibinfo {author} {\bibfnamefont {H.~J.}\ \bibnamefont
  {Kimble}},\ }\bibfield  {title} {\enquote {\bibinfo {title} {{The quantum
  internet}},}\ }\href {\doibase 10.1038/nature07127} {\bibfield  {journal}
  {\bibinfo  {journal} {Nature}\ }\textbf {\bibinfo {volume} {453}},\ \bibinfo
  {pages} {1023--1030} (\bibinfo {year} {2008})},\ \Eprint
  {http://arxiv.org/abs/0806.4195} {arXiv:0806.4195} \BibitemShut {NoStop}%
\bibitem [{\citenamefont {Wehner}\ \emph {et~al.}(2018)\citenamefont {Wehner},
  \citenamefont {Elkouss},\ and\ \citenamefont {Hanson}}]{Wehner2018}%
  \BibitemOpen
  \bibfield  {author} {\bibinfo {author} {\bibfnamefont {Stephanie}\
  \bibnamefont {Wehner}}, \bibinfo {author} {\bibfnamefont {David}\
  \bibnamefont {Elkouss}}, \ and\ \bibinfo {author} {\bibfnamefont {Ronald}\
  \bibnamefont {Hanson}},\ }\bibfield  {title} {\enquote {\bibinfo {title}
  {{Quantum internet: A vision for the road ahead}},}\ }\href {\doibase
  10.1126/science.aam9288} {\bibfield  {journal} {\bibinfo  {journal}
  {Science}\ }\textbf {\bibinfo {volume} {362}},\ \bibinfo {pages} {eaam9288}
  (\bibinfo {year} {2018})}\BibitemShut {NoStop}%
\bibitem [{\citenamefont {Dunjko}\ and\ \citenamefont
  {Briegel}(2018)}]{Dunjko2017}%
  \BibitemOpen
  \bibfield  {author} {\bibinfo {author} {\bibfnamefont {Vedran}\ \bibnamefont
  {Dunjko}}\ and\ \bibinfo {author} {\bibfnamefont {Hans~J.}\ \bibnamefont
  {Briegel}},\ }\bibfield  {title} {\enquote {\bibinfo {title} {{Machine
  learning {\&} artificial intelligence in the quantum domain: a review of
  recent progress}},}\ }\href {\doibase 10.1088/1361-6633/aab406} {\bibfield
  {journal} {\bibinfo  {journal} {Reports on Progress in Physics}\ }\textbf
  {\bibinfo {volume} {81}},\ \bibinfo {pages} {074001} (\bibinfo {year}
  {2018})},\ \Eprint {http://arxiv.org/abs/1709.02779} {arXiv:1709.02779}
  \BibitemShut {NoStop}%
\bibitem [{\citenamefont {Sasaki}\ and\ \citenamefont
  {Carlini}(2002)}]{Sasaki2002}%
  \BibitemOpen
  \bibfield  {author} {\bibinfo {author} {\bibfnamefont {Masahide}\
  \bibnamefont {Sasaki}}\ and\ \bibinfo {author} {\bibfnamefont {Alberto}\
  \bibnamefont {Carlini}},\ }\bibfield  {title} {\enquote {\bibinfo {title}
  {{Quantum learning and universal quantum matching machine}},}\ }\href
  {\doibase 10.1103/PhysRevA.66.022303} {\bibfield  {journal} {\bibinfo
  {journal} {Physical Review A}\ }\textbf {\bibinfo {volume} {66}},\ \bibinfo
  {pages} {022303} (\bibinfo {year} {2002})}\BibitemShut {NoStop}%
\bibitem [{\citenamefont {Liu}\ and\ \citenamefont
  {Rebentrost}(2018)}]{Liu2017}%
  \BibitemOpen
  \bibfield  {author} {\bibinfo {author} {\bibfnamefont {Nana}\ \bibnamefont
  {Liu}}\ and\ \bibinfo {author} {\bibfnamefont {Patrick}\ \bibnamefont
  {Rebentrost}},\ }\bibfield  {title} {\enquote {\bibinfo {title} {{Quantum
  machine learning for quantum anomaly detection}},}\ }\href {\doibase
  10.1103/PhysRevA.97.042315} {\bibfield  {journal} {\bibinfo  {journal}
  {Physical Review A}\ }\textbf {\bibinfo {volume} {97}},\ \bibinfo {pages}
  {042315} (\bibinfo {year} {2018})},\ \Eprint
  {http://arxiv.org/abs/1710.07405} {arXiv:1710.07405} \BibitemShut {NoStop}%
\bibitem [{\citenamefont {Skotiniotis}\ \emph {et~al.}()\citenamefont
  {Skotiniotis}, \citenamefont {Hotz}, \citenamefont {Calsamiglia},\ and\
  \citenamefont {Mu{\~{n}}oz-Tapia}}]{Skotiniotis2018}%
  \BibitemOpen
  \bibfield  {author} {\bibinfo {author} {\bibfnamefont {M.}~\bibnamefont
  {Skotiniotis}}, \bibinfo {author} {\bibfnamefont {R.}~\bibnamefont {Hotz}},
  \bibinfo {author} {\bibfnamefont {J.}~\bibnamefont {Calsamiglia}}, \ and\
  \bibinfo {author} {\bibfnamefont {R.}~\bibnamefont {Mu{\~{n}}oz-Tapia}},\
  }\bibfield  {title} {\enquote {\bibinfo {title} {{Identification of
  malfunctioning quantum devices}},}\ }\href {http://arxiv.org/abs/1808.02729}
  {\ }\Eprint {http://arxiv.org/abs/1808.02729} {arXiv:1808.02729} \BibitemShut
  {NoStop}%
\bibitem [{\citenamefont {Bisio}\ \emph {et~al.}(2010)\citenamefont {Bisio},
  \citenamefont {Chiribella}, \citenamefont {D'Ariano}, \citenamefont
  {Facchini},\ and\ \citenamefont {Perinotti}}]{Bisio2010}%
  \BibitemOpen
  \bibfield  {author} {\bibinfo {author} {\bibfnamefont {Alessandro}\
  \bibnamefont {Bisio}}, \bibinfo {author} {\bibfnamefont {Giulio}\
  \bibnamefont {Chiribella}}, \bibinfo {author} {\bibfnamefont {Giacomo~Mauro}\
  \bibnamefont {D'Ariano}}, \bibinfo {author} {\bibfnamefont {Stefano}\
  \bibnamefont {Facchini}}, \ and\ \bibinfo {author} {\bibfnamefont {Paolo}\
  \bibnamefont {Perinotti}},\ }\bibfield  {title} {\enquote {\bibinfo {title}
  {{Optimal quantum learning of a unitary transformation}},}\ }\href {\doibase
  10.1103/PhysRevA.81.032324} {\bibfield  {journal} {\bibinfo  {journal}
  {Physical Review A}\ }\textbf {\bibinfo {volume} {81}},\ \bibinfo {pages}
  {032324} (\bibinfo {year} {2010})}\BibitemShut {NoStop}%
\bibitem [{\citenamefont {Bisio}\ \emph {et~al.}(2011)\citenamefont {Bisio},
  \citenamefont {DʼAriano}, \citenamefont {Perinotti},\ and\ \citenamefont
  {Sedl{\'{a}}k}}]{Bisio2011a}%
  \BibitemOpen
  \bibfield  {author} {\bibinfo {author} {\bibfnamefont {Alessandro}\
  \bibnamefont {Bisio}}, \bibinfo {author} {\bibfnamefont {Giacomo~Mauro}\
  \bibnamefont {DʼAriano}}, \bibinfo {author} {\bibfnamefont {Paolo}\
  \bibnamefont {Perinotti}}, \ and\ \bibinfo {author} {\bibfnamefont {Michal}\
  \bibnamefont {Sedl{\'{a}}k}},\ }\bibfield  {title} {\enquote {\bibinfo
  {title} {{Quantum learning algorithms for quantum measurements}},}\ }\href
  {\doibase 10.1016/j.physleta.2011.08.002} {\bibfield  {journal} {\bibinfo
  {journal} {Physics Letters A}\ }\textbf {\bibinfo {volume} {375}},\ \bibinfo
  {pages} {3425--3434} (\bibinfo {year} {2011})},\ \Eprint
  {http://arxiv.org/abs/1103.0480} {arXiv:1103.0480} \BibitemShut {NoStop}%
\bibitem [{\citenamefont {Guta}\ and\ \citenamefont
  {Kotlowski}(2010)}]{Guta2010}%
  \BibitemOpen
  \bibfield  {author} {\bibinfo {author} {\bibfnamefont {Madalin}\ \bibnamefont
  {Guta}}\ and\ \bibinfo {author} {\bibfnamefont {Wojciech}\ \bibnamefont
  {Kotlowski}},\ }\bibfield  {title} {\enquote {\bibinfo {title} {{Quantum
  learning: asymptotically optimal classification of qubit states}},}\ }\href
  {\doibase 10.1088/1367-2630/12/12/123032} {\bibfield  {journal} {\bibinfo
  {journal} {New Journal of Physics}\ }\textbf {\bibinfo {volume} {12}},\
  \bibinfo {pages} {123032} (\bibinfo {year} {2010})},\ \Eprint
  {http://arxiv.org/abs/1004.2468} {arXiv:1004.2468} \BibitemShut {NoStop}%
\bibitem [{\citenamefont {Sent{\'{i}}s}\ \emph {et~al.}(2012)\citenamefont
  {Sent{\'{i}}s}, \citenamefont {Calsamiglia}, \citenamefont
  {Mu{\~{n}}oz-Tapia},\ and\ \citenamefont {Bagan}}]{Sentis2012a}%
  \BibitemOpen
  \bibfield  {author} {\bibinfo {author} {\bibfnamefont {Gael}\ \bibnamefont
  {Sent{\'{i}}s}}, \bibinfo {author} {\bibfnamefont {John}\ \bibnamefont
  {Calsamiglia}}, \bibinfo {author} {\bibfnamefont {Ramon}\ \bibnamefont
  {Mu{\~{n}}oz-Tapia}}, \ and\ \bibinfo {author} {\bibfnamefont {Emilio}\
  \bibnamefont {Bagan}},\ }\bibfield  {title} {\enquote {\bibinfo {title}
  {{Quantum learning without quantum memory}},}\ }\href {\doibase
  10.1038/srep00708} {\bibfield  {journal} {\bibinfo  {journal} {Scientific
  Reports}\ }\textbf {\bibinfo {volume} {2}},\ \bibinfo {pages} {708} (\bibinfo
  {year} {2012})}\BibitemShut {NoStop}%
\bibitem [{\citenamefont {Sent{\'{i}}s}\ \emph {et~al.}(2015)\citenamefont
  {Sent{\'{i}}s}, \citenamefont {Guta},\ and\ \citenamefont
  {Adesso}}]{Sentis2014a}%
  \BibitemOpen
  \bibfield  {author} {\bibinfo {author} {\bibfnamefont {Gael}\ \bibnamefont
  {Sent{\'{i}}s}}, \bibinfo {author} {\bibfnamefont {Madalin}\ \bibnamefont
  {Guta}}, \ and\ \bibinfo {author} {\bibfnamefont {Gerardo}\ \bibnamefont
  {Adesso}},\ }\bibfield  {title} {\enquote {\bibinfo {title} {{Quantum
  learning of coherent states}},}\ }\href {\doibase
  10.1140/epjqt/s40507-015-0030-4} {\bibfield  {journal} {\bibinfo  {journal}
  {EPJ Quantum Technology}\ }\textbf {\bibinfo {volume} {2}},\ \bibinfo {pages}
  {17} (\bibinfo {year} {2015})},\ \Eprint {http://arxiv.org/abs/1410.8700}
  {arXiv:1410.8700} \BibitemShut {NoStop}%
\bibitem [{\citenamefont {Fanizza}\ \emph {et~al.}(2019)\citenamefont
  {Fanizza}, \citenamefont {Mari},\ and\ \citenamefont
  {Giovannetti}}]{Fanizza2018}%
  \BibitemOpen
  \bibfield  {author} {\bibinfo {author} {\bibfnamefont {Marco}\ \bibnamefont
  {Fanizza}}, \bibinfo {author} {\bibfnamefont {Andrea}\ \bibnamefont {Mari}},
  \ and\ \bibinfo {author} {\bibfnamefont {Vittorio}\ \bibnamefont
  {Giovannetti}},\ }\bibfield  {title} {\enquote {\bibinfo {title} {{Optimal
  universal learning machines for quantum state discrimination}},}\ }\href
  {\doibase 10.1109/TIT.2019.2916646} {\bibfield  {journal} {\bibinfo
  {journal} {IEEE Transactions on Information Theory}\ ,\ \bibinfo {pages}
  {1--1}} (\bibinfo {year} {2019})},\ \Eprint {http://arxiv.org/abs/1805.03477}
  {arXiv:1805.03477} \BibitemShut {NoStop}%
\bibitem [{\citenamefont {Hastie}\ \emph {et~al.}(2001)\citenamefont {Hastie},
  \citenamefont {Tibshirani},\ and\ \citenamefont {Friedman}}]{Hastie2001}%
  \BibitemOpen
  \bibfield  {author} {\bibinfo {author} {\bibfnamefont {Trevor}\ \bibnamefont
  {Hastie}}, \bibinfo {author} {\bibfnamefont {Robert}\ \bibnamefont
  {Tibshirani}}, \ and\ \bibinfo {author} {\bibfnamefont {Jerome}\ \bibnamefont
  {Friedman}},\ }\href@noop {} {\enquote {\bibinfo {title} {{The elements of
  statistical learning: data mining, inference, and prediction}},}\ } (\bibinfo
  {year} {2001})\BibitemShut {NoStop}%
\bibitem [{\citenamefont {Devroye}\ \emph {et~al.}(1996)\citenamefont
  {Devroye}, \citenamefont {Gy{\"{o}}rfi},\ and\ \citenamefont
  {Lugosi}}]{Devroye2013}%
  \BibitemOpen
  \bibfield  {author} {\bibinfo {author} {\bibfnamefont {Luc}\ \bibnamefont
  {Devroye}}, \bibinfo {author} {\bibfnamefont {L{\'{a}}szl{\'{o}}}\
  \bibnamefont {Gy{\"{o}}rfi}}, \ and\ \bibinfo {author} {\bibfnamefont
  {G{\'{a}}bor}\ \bibnamefont {Lugosi}},\ }\href@noop {} {\emph {\bibinfo
  {title} {{A probabilistic theory of pattern recognition}}}}\ (\bibinfo
  {publisher} {Springer Science {\&} Business Media},\ \bibinfo {year}
  {1996})\BibitemShut {NoStop}%
\bibitem [{\citenamefont {Monr{\`{a}}s}\ \emph {et~al.}(2017)\citenamefont
  {Monr{\`{a}}s}, \citenamefont {Sent{\'{i}}s},\ and\ \citenamefont
  {Wittek}}]{Monras2017}%
  \BibitemOpen
  \bibfield  {author} {\bibinfo {author} {\bibfnamefont {Alex}\ \bibnamefont
  {Monr{\`{a}}s}}, \bibinfo {author} {\bibfnamefont {Gael}\ \bibnamefont
  {Sent{\'{i}}s}}, \ and\ \bibinfo {author} {\bibfnamefont {Peter}\
  \bibnamefont {Wittek}},\ }\bibfield  {title} {\enquote {\bibinfo {title}
  {{Inductive Supervised Quantum Learning}},}\ }\href {\doibase
  10.1103/PhysRevLett.118.190503} {\bibfield  {journal} {\bibinfo  {journal}
  {Physical Review Letters}\ }\textbf {\bibinfo {volume} {118}},\ \bibinfo
  {pages} {190503} (\bibinfo {year} {2017})},\ \Eprint
  {http://arxiv.org/abs/1605.07541} {arXiv:1605.07541} \BibitemShut {NoStop}%
\bibitem [{\citenamefont {Dunjko}\ \emph {et~al.}(2016)\citenamefont {Dunjko},
  \citenamefont {Taylor},\ and\ \citenamefont {Briegel}}]{Dunjko2016}%
  \BibitemOpen
  \bibfield  {author} {\bibinfo {author} {\bibfnamefont {Vedran}\ \bibnamefont
  {Dunjko}}, \bibinfo {author} {\bibfnamefont {Jacob~M.}\ \bibnamefont
  {Taylor}}, \ and\ \bibinfo {author} {\bibfnamefont {Hans~J.}\ \bibnamefont
  {Briegel}},\ }\bibfield  {title} {\enquote {\bibinfo {title}
  {{Quantum-Enhanced Machine Learning}},}\ }\href {\doibase
  10.1103/PhysRevLett.117.130501} {\bibfield  {journal} {\bibinfo  {journal}
  {Physical Review Letters}\ }\textbf {\bibinfo {volume} {117}},\ \bibinfo
  {pages} {130501} (\bibinfo {year} {2016})},\ \Eprint
  {http://arxiv.org/abs/1610.08251} {arXiv:1610.08251} \BibitemShut {NoStop}%
\bibitem [{\citenamefont {Aloise}\ \emph {et~al.}(2009)\citenamefont {Aloise},
  \citenamefont {Deshpande}, \citenamefont {Hansen},\ and\ \citenamefont
  {Popat}}]{Aloise2009}%
  \BibitemOpen
  \bibfield  {author} {\bibinfo {author} {\bibfnamefont {Daniel}\ \bibnamefont
  {Aloise}}, \bibinfo {author} {\bibfnamefont {Amit}\ \bibnamefont
  {Deshpande}}, \bibinfo {author} {\bibfnamefont {Pierre}\ \bibnamefont
  {Hansen}}, \ and\ \bibinfo {author} {\bibfnamefont {Preyas}\ \bibnamefont
  {Popat}},\ }\bibfield  {title} {\enquote {\bibinfo {title} {{NP-hardness of
  Euclidean sum-of-squares clustering}},}\ }\href {\doibase
  10.1007/s10994-009-5103-0} {\bibfield  {journal} {\bibinfo  {journal}
  {Machine Learning}\ }\textbf {\bibinfo {volume} {75}},\ \bibinfo {pages}
  {245--248} (\bibinfo {year} {2009})}\BibitemShut {NoStop}%
\bibitem [{\citenamefont {Ben-David}()}]{Ben-David2015}%
  \BibitemOpen
  \bibfield  {author} {\bibinfo {author} {\bibfnamefont {Shai}\ \bibnamefont
  {Ben-David}},\ }\bibfield  {title} {\enquote {\bibinfo {title} {{Clustering
  is Easy When ....What?}}}\ }\href {http://arxiv.org/abs/1510.05336} {\
  }\Eprint {http://arxiv.org/abs/1510.05336} {arXiv:1510.05336} \BibitemShut
  {NoStop}%
\bibitem [{\citenamefont {Bu{\v{z}}ek}\ \emph {et~al.}(2006)\citenamefont
  {Bu{\v{z}}ek}, \citenamefont {Hillery}, \citenamefont {Ziman},\ and\
  \citenamefont {Ro{\v{s}}ko}}]{Buzek2006}%
  \BibitemOpen
  \bibfield  {author} {\bibinfo {author} {\bibfnamefont {Vladim{\'{i}}r}\
  \bibnamefont {Bu{\v{z}}ek}}, \bibinfo {author} {\bibfnamefont {Mark}\
  \bibnamefont {Hillery}}, \bibinfo {author} {\bibfnamefont {M{\'{a}}rio}\
  \bibnamefont {Ziman}}, \ and\ \bibinfo {author} {\bibfnamefont
  {Mari{\'{a}}n}\ \bibnamefont {Ro{\v{s}}ko}},\ }\bibfield  {title} {\enquote
  {\bibinfo {title} {{Programmable Quantum Processors}},}\ }\href {\doibase
  10.1007/s11128-006-0028-z} {\bibfield  {journal} {\bibinfo  {journal}
  {Quantum Information Processing}\ }\textbf {\bibinfo {volume} {5}},\ \bibinfo
  {pages} {313--420} (\bibinfo {year} {2006})}\BibitemShut {NoStop}%
\bibitem [{\citenamefont {Helstrom}(1976)}]{Helstrom1976}%
  \BibitemOpen
  \bibfield  {author} {\bibinfo {author} {\bibfnamefont {C.~W.}\ \bibnamefont
  {Helstrom}},\ }\href@noop {} {\emph {\bibinfo {title} {{Quantum Detection and
  Estimation Theory}}}}\ (\bibinfo  {publisher} {Academic Press},\ \bibinfo
  {address} {New York},\ \bibinfo {year} {1976})\BibitemShut {NoStop}%
\bibitem [{\citenamefont {Barnett}(2001)}]{Barnett2001}%
  \BibitemOpen
  \bibfield  {author} {\bibinfo {author} {\bibfnamefont {Stephen~M.}\
  \bibnamefont {Barnett}},\ }\bibfield  {title} {\enquote {\bibinfo {title}
  {{Minimum-error discrimination between multiply symmetric states}},}\ }\href
  {\doibase 10.1103/PhysRevA.64.030303} {\bibfield  {journal} {\bibinfo
  {journal} {Physical Review A}\ }\textbf {\bibinfo {volume} {64}},\ \bibinfo
  {pages} {030303(R)} (\bibinfo {year} {2001})}\BibitemShut {NoStop}%
\bibitem [{\citenamefont {Chiribella}\ \emph {et~al.}(2004)\citenamefont
  {Chiribella}, \citenamefont {D'Ariano}, \citenamefont {Perinotti},\ and\
  \citenamefont {Sacchi}}]{Chiribella2004}%
  \BibitemOpen
  \bibfield  {author} {\bibinfo {author} {\bibfnamefont {Giulio}\ \bibnamefont
  {Chiribella}}, \bibinfo {author} {\bibfnamefont {Giacomo~Mauro}\ \bibnamefont
  {D'Ariano}}, \bibinfo {author} {\bibfnamefont {Paolo}\ \bibnamefont
  {Perinotti}}, \ and\ \bibinfo {author} {\bibfnamefont {Massimiliano~F.}\
  \bibnamefont {Sacchi}},\ }\bibfield  {title} {\enquote {\bibinfo {title}
  {{Covariant quantum measurements that maximize the likelihood}},}\ }\href
  {\doibase 10.1103/PhysRevA.70.062105} {\bibfield  {journal} {\bibinfo
  {journal} {Physical Review A}\ }\textbf {\bibinfo {volume} {70}},\ \bibinfo
  {pages} {062105} (\bibinfo {year} {2004})},\ \Eprint
  {http://arxiv.org/abs/0403083} {arXiv:0403083 [quant-ph]} \BibitemShut
  {NoStop}%
\bibitem [{\citenamefont {Chiribella}\ \emph {et~al.}(2006)\citenamefont
  {Chiribella}, \citenamefont {D'Ariano}, \citenamefont {Perinotti},\ and\
  \citenamefont {Sacchi}}]{Chiribella2006a}%
  \BibitemOpen
  \bibfield  {author} {\bibinfo {author} {\bibfnamefont {Giulio}\ \bibnamefont
  {Chiribella}}, \bibinfo {author} {\bibfnamefont {Giacomo~Mauro}\ \bibnamefont
  {D'Ariano}}, \bibinfo {author} {\bibfnamefont {Paolo}\ \bibnamefont
  {Perinotti}}, \ and\ \bibinfo {author} {\bibfnamefont {Massimiliano~F.}\
  \bibnamefont {Sacchi}},\ }\bibfield  {title} {\enquote {\bibinfo {title}
  {{Maximum likelihood estimation for a group of physical transformations}},}\
  }\href {\doibase 10.1142/S0219749906002018} {\bibfield  {journal} {\bibinfo
  {journal} {International Journal of Quantum Information}\ }\textbf {\bibinfo
  {volume} {04}},\ \bibinfo {pages} {453--472} (\bibinfo {year}
  {2006})}\BibitemShut {NoStop}%
\bibitem [{\citenamefont {Audenaert}\ \emph {et~al.}(2007)\citenamefont
  {Audenaert}, \citenamefont {Calsamiglia}, \citenamefont {Munoz-Tapia},
  \citenamefont {Bagan}, \citenamefont {Masanes}, \citenamefont {Acin},\ and\
  \citenamefont {Verstraete}}]{Audenaert2007}%
  \BibitemOpen
  \bibfield  {author} {\bibinfo {author} {\bibfnamefont {K.~M.~R.}\
  \bibnamefont {Audenaert}}, \bibinfo {author} {\bibfnamefont {J.}~\bibnamefont
  {Calsamiglia}}, \bibinfo {author} {\bibfnamefont {R.}~\bibnamefont
  {Munoz-Tapia}}, \bibinfo {author} {\bibfnamefont {E.}~\bibnamefont {Bagan}},
  \bibinfo {author} {\bibfnamefont {Ll.}\ \bibnamefont {Masanes}}, \bibinfo
  {author} {\bibfnamefont {A.}~\bibnamefont {Acin}}, \ and\ \bibinfo {author}
  {\bibfnamefont {F.}~\bibnamefont {Verstraete}},\ }\bibfield  {title}
  {\enquote {\bibinfo {title} {{Discriminating States: The Quantum Chernoff
  Bound}},}\ }\href {\doibase 10.1103/PhysRevLett.98.160501} {\bibfield
  {journal} {\bibinfo  {journal} {Physical Review Letters}\ }\textbf {\bibinfo
  {volume} {98}},\ \bibinfo {pages} {160501} (\bibinfo {year}
  {2007})}\BibitemShut {NoStop}%
\bibitem [{\citenamefont {Krovi}\ \emph {et~al.}(2015)\citenamefont {Krovi},
  \citenamefont {Guha}, \citenamefont {Dutton},\ and\ \citenamefont
  {da~Silva}}]{Krovi2015a}%
  \BibitemOpen
  \bibfield  {author} {\bibinfo {author} {\bibfnamefont {Hari}\ \bibnamefont
  {Krovi}}, \bibinfo {author} {\bibfnamefont {Saikat}\ \bibnamefont {Guha}},
  \bibinfo {author} {\bibfnamefont {Zachary}\ \bibnamefont {Dutton}}, \ and\
  \bibinfo {author} {\bibfnamefont {Marcus~P.}\ \bibnamefont {da~Silva}},\
  }\bibfield  {title} {\enquote {\bibinfo {title} {{Optimal measurements for
  symmetric quantum states with applications to optical communication}},}\
  }\href {\doibase 10.1103/PhysRevA.92.062333} {\bibfield  {journal} {\bibinfo
  {journal} {Physical Review A}\ }\textbf {\bibinfo {volume} {92}},\ \bibinfo
  {pages} {062333} (\bibinfo {year} {2015})},\ \Eprint
  {http://arxiv.org/abs/1507.04737} {arXiv:1507.04737} \BibitemShut {NoStop}%
\bibitem [{\citenamefont {Knuth}(1998)}]{Knuth1998}%
  \BibitemOpen
  \bibfield  {author} {\bibinfo {author} {\bibfnamefont {Donald~E.}\
  \bibnamefont {Knuth}},\ }\href@noop {} {\emph {\bibinfo {title} {{The Art of
  Computer Programming Vol. 3: Sorting and Searching}}}}\ (\bibinfo
  {publisher} {Addison-Wesley},\ \bibinfo {year} {1998})\BibitemShut {NoStop}%
\bibitem [{\citenamefont {Abramowitz}\ and\ \citenamefont
  {Stegun}(1965)}]{Abramowitz1965}%
  \BibitemOpen
  \bibfield  {author} {\bibinfo {author} {\bibfnamefont {M.}~\bibnamefont
  {Abramowitz}}\ and\ \bibinfo {author} {\bibfnamefont {I.~A.}\ \bibnamefont
  {Stegun}},\ }\href@noop {} {\emph {\bibinfo {title} {{Handbook of
  Mathematical Functions: with Formulas, Graphs, and Mathematical Tables}}}}\
  (\bibinfo  {publisher} {Dover Publications},\ \bibinfo {address} {New York},\
  \bibinfo {year} {1965})\BibitemShut {NoStop}%
\bibitem [{\citenamefont {Korf}(1998)}]{Korf1998}%
  \BibitemOpen
  \bibfield  {author} {\bibinfo {author} {\bibfnamefont {Richard~E}\
  \bibnamefont {Korf}},\ }\bibfield  {title} {\enquote {\bibinfo {title} {{A
  complete anytime algorithm for number partitioning}},}\ }\href@noop {}
  {\bibfield  {journal} {\bibinfo  {journal} {Artificial Intelligence}\
  }\textbf {\bibinfo {volume} {106}},\ \bibinfo {pages} {181--203} (\bibinfo
  {year} {1998})}\BibitemShut {NoStop}%
\bibitem [{\citenamefont {Holevo}(1982)}]{Holevo1982}%
  \BibitemOpen
  \bibfield  {author} {\bibinfo {author} {\bibfnamefont {Alexander~S.}\
  \bibnamefont {Holevo}},\ }\href {\doibase 10.1007/978-88-7642-378-9} {\emph
  {\bibinfo {title} {{Probabilistic and Statistical Aspects of Quantum
  Theory}}}}\ (\bibinfo  {publisher} {North-Holland},\ \bibinfo {address}
  {Amsterdam},\ \bibinfo {year} {1982})\BibitemShut {NoStop}%
\bibitem [{\citenamefont {Holevo}(1973)}]{Holevo1973a}%
  \BibitemOpen
  \bibfield  {author} {\bibinfo {author} {\bibfnamefont {A.S}\ \bibnamefont
  {Holevo}},\ }\bibfield  {title} {\enquote {\bibinfo {title} {{Statistical
  decision theory for quantum systems}},}\ }\href {\doibase
  10.1016/0047-259X(73)90028-6} {\bibfield  {journal} {\bibinfo  {journal}
  {Journal of Multivariate Analysis}\ }\textbf {\bibinfo {volume} {3}},\
  \bibinfo {pages} {337--394} (\bibinfo {year} {1973})}\BibitemShut {NoStop}%
\bibitem [{\citenamefont {Yuen}\ \emph {et~al.}(1975)\citenamefont {Yuen},
  \citenamefont {Kennedy},\ and\ \citenamefont {Lax}}]{Yuen1975}%
  \BibitemOpen
  \bibfield  {author} {\bibinfo {author} {\bibfnamefont {H.}~\bibnamefont
  {Yuen}}, \bibinfo {author} {\bibfnamefont {R.}~\bibnamefont {Kennedy}}, \
  and\ \bibinfo {author} {\bibfnamefont {M.}~\bibnamefont {Lax}},\ }\bibfield
  {title} {\enquote {\bibinfo {title} {{Optimum testing of multiple hypotheses
  in quantum detection theory}},}\ }\href {\doibase 10.1109/TIT.1975.1055351}
  {\bibfield  {journal} {\bibinfo  {journal} {IEEE Transactions on Information
  Theory}\ }\textbf {\bibinfo {volume} {21}},\ \bibinfo {pages} {125--134}
  (\bibinfo {year} {1975})}\BibitemShut {NoStop}%
\bibitem [{\citenamefont {{Dalla Pozza}}\ and\ \citenamefont
  {Pierobon}(2015)}]{DallaPozza2015}%
  \BibitemOpen
  \bibfield  {author} {\bibinfo {author} {\bibfnamefont {Nicola}\ \bibnamefont
  {{Dalla Pozza}}}\ and\ \bibinfo {author} {\bibfnamefont {Gianfranco}\
  \bibnamefont {Pierobon}},\ }\bibfield  {title} {\enquote {\bibinfo {title}
  {{Optimality of square-root measurements in quantum state discrimination}},}\
  }\href {\doibase 10.1103/PhysRevA.91.042334} {\bibfield  {journal} {\bibinfo
  {journal} {Physical Review A}\ }\textbf {\bibinfo {volume} {91}},\ \bibinfo
  {pages} {042334} (\bibinfo {year} {2015})},\ \Eprint
  {http://arxiv.org/abs/1504.04908} {arXiv:1504.04908} \BibitemShut {NoStop}%
\bibitem [{\citenamefont {Sent{\'{i}}s}\ \emph {et~al.}(2016)\citenamefont
  {Sent{\'{i}}s}, \citenamefont {Bagan}, \citenamefont {Calsamiglia},
  \citenamefont {Chiribella},\ and\ \citenamefont
  {Mu{\~{n}}oz-Tapia}}]{Sentis2016}%
  \BibitemOpen
  \bibfield  {author} {\bibinfo {author} {\bibfnamefont {Gael}\ \bibnamefont
  {Sent{\'{i}}s}}, \bibinfo {author} {\bibfnamefont {Emilio}\ \bibnamefont
  {Bagan}}, \bibinfo {author} {\bibfnamefont {John}\ \bibnamefont
  {Calsamiglia}}, \bibinfo {author} {\bibfnamefont {Giulio}\ \bibnamefont
  {Chiribella}}, \ and\ \bibinfo {author} {\bibfnamefont {Ramon}\ \bibnamefont
  {Mu{\~{n}}oz-Tapia}},\ }\bibfield  {title} {\enquote {\bibinfo {title}
  {{Quantum Change Point}},}\ }\href {\doibase 10.1103/PhysRevLett.117.150502}
  {\bibfield  {journal} {\bibinfo  {journal} {Physical Review Letters}\
  }\textbf {\bibinfo {volume} {117}},\ \bibinfo {pages} {150502} (\bibinfo
  {year} {2016})},\ \Eprint {http://arxiv.org/abs/1605.01916}
  {arXiv:1605.01916} \BibitemShut {NoStop}%
\bibitem [{\citenamefont {Horn}\ and\ \citenamefont
  {Johnson}(2013)}]{Horn2013}%
  \BibitemOpen
  \bibfield  {author} {\bibinfo {author} {\bibfnamefont {Roger~A.}\
  \bibnamefont {Horn}}\ and\ \bibinfo {author} {\bibfnamefont {Charles~R.}\
  \bibnamefont {Johnson}},\ }\href {www.cambridge.org/9780521548236} {\emph
  {\bibinfo {title} {{Matrix analysis}}}},\ \bibinfo {edition} {2nd}\ ed.\
  (\bibinfo  {publisher} {Cambridge University Press},\ \bibinfo {address}
  {Cambridge, England},\ \bibinfo {year} {2013})\BibitemShut {NoStop}%
\bibitem [{\citenamefont {Ivanovic}(1987)}]{Ivanovic1987}%
  \BibitemOpen
  \bibfield  {author} {\bibinfo {author} {\bibfnamefont {I.D.}\ \bibnamefont
  {Ivanovic}},\ }\bibfield  {title} {\enquote {\bibinfo {title} {{How to
  differentiate between non-orthogonal states}},}\ }\href {\doibase
  10.1016/0375-9601(87)90222-2} {\bibfield  {journal} {\bibinfo  {journal}
  {Physics Letters A}\ }\textbf {\bibinfo {volume} {123}},\ \bibinfo {pages}
  {257--259} (\bibinfo {year} {1987})}\BibitemShut {NoStop}%
\bibitem [{\citenamefont {Dieks}(1988)}]{Dieks1988}%
  \BibitemOpen
  \bibfield  {author} {\bibinfo {author} {\bibfnamefont {D.}~\bibnamefont
  {Dieks}},\ }\bibfield  {title} {\enquote {\bibinfo {title} {{Overlap and
  distinguishability of quantum states}},}\ }\href {\doibase
  10.1016/0375-9601(88)90840-7} {\bibfield  {journal} {\bibinfo  {journal}
  {Physics Letters A}\ }\textbf {\bibinfo {volume} {126}},\ \bibinfo {pages}
  {303--306} (\bibinfo {year} {1988})}\BibitemShut {NoStop}%
\bibitem [{\citenamefont {Peres}(1988)}]{Peres1988}%
  \BibitemOpen
  \bibfield  {author} {\bibinfo {author} {\bibfnamefont {Asher}\ \bibnamefont
  {Peres}},\ }\bibfield  {title} {\enquote {\bibinfo {title} {{How to
  differentiate between non-orthogonal states}},}\ }\href {\doibase
  10.1016/0375-9601(88)91034-1} {\bibfield  {journal} {\bibinfo  {journal}
  {Physics Letters A}\ }\textbf {\bibinfo {volume} {128}},\ \bibinfo {pages}
  {19} (\bibinfo {year} {1988})}\BibitemShut {NoStop}%
\bibitem [{\citenamefont {Chefles}\ and\ \citenamefont
  {Barnett}(1998)}]{Chefles1998a}%
  \BibitemOpen
  \bibfield  {author} {\bibinfo {author} {\bibfnamefont {Anthony}\ \bibnamefont
  {Chefles}}\ and\ \bibinfo {author} {\bibfnamefont {Stephen~M}\ \bibnamefont
  {Barnett}},\ }\bibfield  {title} {\enquote {\bibinfo {title} {{Optimum
  unambiguous discrimination between linearly independent symmetric states}},}\
  }\href {\doibase 10.1016/S0375-9601(98)00827-5} {\bibfield  {journal}
  {\bibinfo  {journal} {Physics Letters A}\ }\textbf {\bibinfo {volume}
  {250}},\ \bibinfo {pages} {223--229} (\bibinfo {year} {1998})}\BibitemShut
  {NoStop}%
\bibitem [{\citenamefont {Lloyd}(1982)}]{Lloyd1982}%
  \BibitemOpen
  \bibfield  {author} {\bibinfo {author} {\bibfnamefont {S.}~\bibnamefont
  {Lloyd}},\ }\bibfield  {title} {\enquote {\bibinfo {title} {{Least squares
  quantization in PCM}},}\ }\href {\doibase 10.1109/TIT.1982.1056489}
  {\bibfield  {journal} {\bibinfo  {journal} {IEEE Transactions on Information
  Theory}\ }\textbf {\bibinfo {volume} {28}},\ \bibinfo {pages} {129--137}
  (\bibinfo {year} {1982})}\BibitemShut {NoStop}%
\bibitem [{\citenamefont {Harrow}(2005)}]{Harrow2005}%
  \BibitemOpen
  \bibfield  {author} {\bibinfo {author} {\bibfnamefont {Aram~W}\ \bibnamefont
  {Harrow}},\ }\emph {\bibinfo {title} {{Applications of coherent classical
  communication and the Schur transform to quantum information theory}}},\
  \href {http://arxiv.org/abs/quant-ph/0512255} {Ph.D. thesis} (\bibinfo {year}
  {2005}),\ \Eprint {http://arxiv.org/abs/0512255} {arXiv:0512255 [quant-ph]}
  \BibitemShut {NoStop}%
\bibitem [{\citenamefont {Krovi}(2019)}]{Krovi2018}%
  \BibitemOpen
  \bibfield  {author} {\bibinfo {author} {\bibfnamefont {Hari}\ \bibnamefont
  {Krovi}},\ }\bibfield  {title} {\enquote {\bibinfo {title} {{An efficient
  high dimensional quantum Schur transform}},}\ }\href {\doibase
  10.22331/q-2019-02-14-122} {\bibfield  {journal} {\bibinfo  {journal}
  {Quantum}\ }\textbf {\bibinfo {volume} {3}},\ \bibinfo {pages} {122}
  (\bibinfo {year} {2019})},\ \Eprint {http://arxiv.org/abs/1804.00055}
  {arXiv:1804.00055} \BibitemShut {NoStop}%
\bibitem [{\citenamefont {Sent{\'{i}}s}\ \emph {et~al.}(2017)\citenamefont
  {Sent{\'{i}}s}, \citenamefont {Calsamiglia},\ and\ \citenamefont
  {Mu{\~{n}}oz-Tapia}}]{Sentis2017}%
  \BibitemOpen
  \bibfield  {author} {\bibinfo {author} {\bibfnamefont {Gael}\ \bibnamefont
  {Sent{\'{i}}s}}, \bibinfo {author} {\bibfnamefont {John}\ \bibnamefont
  {Calsamiglia}}, \ and\ \bibinfo {author} {\bibfnamefont {Ramon}\ \bibnamefont
  {Mu{\~{n}}oz-Tapia}},\ }\bibfield  {title} {\enquote {\bibinfo {title}
  {{Exact Identification of a Quantum Change Point}},}\ }\href {\doibase
  10.1103/PhysRevLett.119.140506} {\bibfield  {journal} {\bibinfo  {journal}
  {Physical Review Letters}\ }\textbf {\bibinfo {volume} {119}},\ \bibinfo
  {pages} {140506} (\bibinfo {year} {2017})},\ \Eprint
  {http://arxiv.org/abs/1707.07769} {arXiv:1707.07769} \BibitemShut {NoStop}%
\bibitem [{\citenamefont {Korff}\ and\ \citenamefont
  {Kempe}(2004)}]{Korff2004}%
  \BibitemOpen
  \bibfield  {author} {\bibinfo {author} {\bibfnamefont {Joshua~Von}\
  \bibnamefont {Korff}}\ and\ \bibinfo {author} {\bibfnamefont {Julia}\
  \bibnamefont {Kempe}},\ }\bibfield  {title} {\enquote {\bibinfo {title}
  {{Quantum Advantage in Transmitting a Permutation}},}\ }\href {\doibase
  10.1103/PhysRevLett.93.260502} {\bibfield  {journal} {\bibinfo  {journal}
  {Physical Review Letters}\ }\textbf {\bibinfo {volume} {93}},\ \bibinfo
  {pages} {260502} (\bibinfo {year} {2004})}\BibitemShut {NoStop}%
\bibitem [{\citenamefont {Hillery}\ \emph {et~al.}(2010)\citenamefont
  {Hillery}, \citenamefont {Andersson}, \citenamefont {Barnett},\ and\
  \citenamefont {Oi}}]{Hillery2011}%
  \BibitemOpen
  \bibfield  {author} {\bibinfo {author} {\bibfnamefont {Mark}\ \bibnamefont
  {Hillery}}, \bibinfo {author} {\bibfnamefont {Erika}\ \bibnamefont
  {Andersson}}, \bibinfo {author} {\bibfnamefont {Stephen~M.}\ \bibnamefont
  {Barnett}}, \ and\ \bibinfo {author} {\bibfnamefont {Daniel}\ \bibnamefont
  {Oi}},\ }\bibfield  {title} {\enquote {\bibinfo {title} {{Decision problems
  with quantum black boxes}},}\ }\href {\doibase 10.1080/09500340903203129}
  {\bibfield  {journal} {\bibinfo  {journal} {Journal of Modern Optics}\
  }\textbf {\bibinfo {volume} {57}},\ \bibinfo {pages} {244--252} (\bibinfo
  {year} {2010})},\ \Eprint {http://arxiv.org/abs/1109.4823} {arXiv:1109.4823}
  \BibitemShut {NoStop}%
\bibitem [{\citenamefont {A{\"{i}}meur}\ \emph {et~al.}(2013)\citenamefont
  {A{\"{i}}meur}, \citenamefont {Brassard},\ and\ \citenamefont
  {Gambs}}]{Aimeur2013}%
  \BibitemOpen
  \bibfield  {author} {\bibinfo {author} {\bibfnamefont {Esma}\ \bibnamefont
  {A{\"{i}}meur}}, \bibinfo {author} {\bibfnamefont {Gilles}\ \bibnamefont
  {Brassard}}, \ and\ \bibinfo {author} {\bibfnamefont {S{\'{e}}bastien}\
  \bibnamefont {Gambs}},\ }\bibfield  {title} {\enquote {\bibinfo {title}
  {{Quantum speed-up for unsupervised learning}},}\ }\href {\doibase
  10.1007/s10994-012-5316-5} {\bibfield  {journal} {\bibinfo  {journal}
  {Machine Learning}\ }\textbf {\bibinfo {volume} {90}},\ \bibinfo {pages}
  {261--287} (\bibinfo {year} {2013})}\BibitemShut {NoStop}%
\bibitem [{\citenamefont {Lloyd}\ \emph {et~al.}()\citenamefont {Lloyd},
  \citenamefont {Mohseni},\ and\ \citenamefont {Rebentrost}}]{Lloyd2013}%
  \BibitemOpen
  \bibfield  {author} {\bibinfo {author} {\bibfnamefont {Seth}\ \bibnamefont
  {Lloyd}}, \bibinfo {author} {\bibfnamefont {Masoud}\ \bibnamefont {Mohseni}},
  \ and\ \bibinfo {author} {\bibfnamefont {Patrick}\ \bibnamefont
  {Rebentrost}},\ }\bibfield  {title} {\enquote {\bibinfo {title} {{Quantum
  algorithms for supervised and unsupervised machine learning}},}\ }\href
  {http://arxiv.org/abs/1307.0411} {\ }\Eprint {http://arxiv.org/abs/1307.0411}
  {arXiv:1307.0411} \BibitemShut {NoStop}%
\bibitem [{\citenamefont {Wiebe}\ \emph {et~al.}(2015)\citenamefont {Wiebe},
  \citenamefont {Kapoor},\ and\ \citenamefont {Svore}}]{Wiebe2014a}%
  \BibitemOpen
  \bibfield  {author} {\bibinfo {author} {\bibfnamefont {Nathan}\ \bibnamefont
  {Wiebe}}, \bibinfo {author} {\bibfnamefont {Ashish}\ \bibnamefont {Kapoor}},
  \ and\ \bibinfo {author} {\bibfnamefont {Krysta}\ \bibnamefont {Svore}},\
  }\bibfield  {title} {\enquote {\bibinfo {title} {{Quantum Algorithms for
  Nearest-Neighbor Methods for Supervised and Unsupervised Learning}},}\ }\href
  {\doibase https://doi.org/10.26421/QIC15.3-4} {\bibfield  {journal} {\bibinfo
   {journal} {Quantum Information {\&} Computation}\ }\textbf {\bibinfo
  {volume} {15}},\ \bibinfo {pages} {0318--0358} (\bibinfo {year} {2015})},\
  \Eprint {http://arxiv.org/abs/1401.2142} {arXiv:1401.2142} \BibitemShut
  {NoStop}%
\bibitem [{\citenamefont {Kerenidis}\ \emph {et~al.}()\citenamefont
  {Kerenidis}, \citenamefont {Landman}, \citenamefont {Luongo},\ and\
  \citenamefont {Prakash}}]{Kerenidis2018}%
  \BibitemOpen
  \bibfield  {author} {\bibinfo {author} {\bibfnamefont {Iordanis}\
  \bibnamefont {Kerenidis}}, \bibinfo {author} {\bibfnamefont {Jonas}\
  \bibnamefont {Landman}}, \bibinfo {author} {\bibfnamefont {Alessandro}\
  \bibnamefont {Luongo}}, \ and\ \bibinfo {author} {\bibfnamefont {Anupam}\
  \bibnamefont {Prakash}},\ }\bibfield  {title} {\enquote {\bibinfo {title}
  {{q-means: A quantum algorithm for unsupervised machine learning}},}\ }\href
  {http://arxiv.org/abs/1812.03584} {\ }\Eprint
  {http://arxiv.org/abs/1812.03584} {arXiv:1812.03584} \BibitemShut {NoStop}%
\bibitem [{\citenamefont {Giovannetti}\ \emph {et~al.}(2008)\citenamefont
  {Giovannetti}, \citenamefont {Lloyd},\ and\ \citenamefont
  {Maccone}}]{Giovannetti2008}%
  \BibitemOpen
  \bibfield  {author} {\bibinfo {author} {\bibfnamefont {Vittorio}\
  \bibnamefont {Giovannetti}}, \bibinfo {author} {\bibfnamefont {Seth}\
  \bibnamefont {Lloyd}}, \ and\ \bibinfo {author} {\bibfnamefont {Lorenzo}\
  \bibnamefont {Maccone}},\ }\bibfield  {title} {\enquote {\bibinfo {title}
  {{Quantum Random Access Memory}},}\ }\href {\doibase
  10.1103/PhysRevLett.100.160501} {\bibfield  {journal} {\bibinfo  {journal}
  {Physical Review Letters}\ }\textbf {\bibinfo {volume} {100}},\ \bibinfo
  {pages} {160501} (\bibinfo {year} {2008})},\ \Eprint
  {http://arxiv.org/abs/0708.1879} {arXiv:0708.1879} \BibitemShut {NoStop}%
\bibitem [{\citenamefont {Chiribella}\ and\ \citenamefont
  {Ebler}(2019)}]{Chiribella2019}%
  \BibitemOpen
  \bibfield  {author} {\bibinfo {author} {\bibfnamefont {Giulio}\ \bibnamefont
  {Chiribella}}\ and\ \bibinfo {author} {\bibfnamefont {Daniel}\ \bibnamefont
  {Ebler}},\ }\bibfield  {title} {\enquote {\bibinfo {title} {{Quantum speedup
  in the identification of cause-effect relations}},}\ }\href {\doibase
  10.1038/s41467-019-09383-8} {\bibfield  {journal} {\bibinfo  {journal}
  {Nature Communications}\ }\textbf {\bibinfo {volume} {10}} (\bibinfo {year}
  {2019}),\ 10.1038/s41467-019-09383-8}\BibitemShut {NoStop}%
\bibitem [{\citenamefont {Flajolet}\ and\ \citenamefont
  {Sedgewick}(2009)}]{Flajolet2009}%
  \BibitemOpen
  \bibfield  {author} {\bibinfo {author} {\bibfnamefont {Philippe}\
  \bibnamefont {Flajolet}}\ and\ \bibinfo {author} {\bibfnamefont {Robert}\
  \bibnamefont {Sedgewick}},\ }\href@noop {} {\emph {\bibinfo {title}
  {{Analytic Combinatorics}}}}\ (\bibinfo  {publisher} {Cambridge University
  Press},\ \bibinfo {year} {2009})\BibitemShut {NoStop}%
\bibitem [{\citenamefont {Andrews}(1976)}]{Andrews1976}%
  \BibitemOpen
  \bibfield  {author} {\bibinfo {author} {\bibfnamefont {George~E.}\
  \bibnamefont {Andrews}},\ }\href@noop {} {\emph {\bibinfo {title}
  {{Encyclopedia of Mathematics and its Applications, Vol. 2, The Theory of
  Partitions}}}}\ (\bibinfo  {publisher} {Addison-Wesley},\ \bibinfo {year}
  {1976})\BibitemShut {NoStop}%
\bibitem [{\citenamefont {Sagan}(2001)}]{Sagan2001}%
  \BibitemOpen
  \bibfield  {author} {\bibinfo {author} {\bibfnamefont {Bruce~E.}\
  \bibnamefont {Sagan}},\ }\href@noop {} {\emph {\bibinfo {title} {{The
  Symmetric Group - Representations, Combinatorial Algorithms, and Symmetric
  Functions}}}}\ (\bibinfo  {publisher} {Springer Science {\&} Business
  Media},\ \bibinfo {address} {New York},\ \bibinfo {year} {2001})\BibitemShut
  {NoStop}%
\bibitem [{\citenamefont {Goodman}\ and\ \citenamefont
  {Wallach}(2009)}]{Goodman2009}%
  \BibitemOpen
  \bibfield  {author} {\bibinfo {author} {\bibfnamefont {Roe}\ \bibnamefont
  {Goodman}}\ and\ \bibinfo {author} {\bibfnamefont {Nolan~R.}\ \bibnamefont
  {Wallach}},\ }\href {\doibase 10.1007/978-0-387-79852-3} {\emph {\bibinfo
  {title} {{Symmetry, Representations, and Invariants}}}},\ edited by\ \bibinfo
  {editor} {\bibfnamefont {S.}~\bibnamefont {Axler}}\ and\ \bibinfo {editor}
  {\bibfnamefont {K.~A.}\ \bibnamefont {Ribet}},\ \bibinfo {series} {Graduate
  Texts in Mathematics}, Vol.\ \bibinfo {volume} {255}\ (\bibinfo  {publisher}
  {Springer New York},\ \bibinfo {address} {New York, NY},\ \bibinfo {year}
  {2009})\BibitemShut {NoStop}%
\bibitem [{\citenamefont {Gliske}\ \emph {et~al.}(2005)\citenamefont {Gliske},
  \citenamefont {Klink},\ and\ \citenamefont {Ton-That}}]{Gliske2005}%
  \BibitemOpen
  \bibfield  {author} {\bibinfo {author} {\bibfnamefont {S.}~\bibnamefont
  {Gliske}}, \bibinfo {author} {\bibfnamefont {W.~H.}\ \bibnamefont {Klink}}, \
  and\ \bibinfo {author} {\bibfnamefont {T.}~\bibnamefont {Ton-That}},\
  }\bibfield  {title} {\enquote {\bibinfo {title} {{Algorithms for Computing
  Generalized U(N) Racah Coefficients}},}\ }\href {\doibase
  10.1007/s10440-005-8345-2} {\bibfield  {journal} {\bibinfo  {journal} {Acta
  Applicandae Mathematicae}\ }\textbf {\bibinfo {volume} {88}},\ \bibinfo
  {pages} {229--249} (\bibinfo {year} {2005})}\BibitemShut {NoStop}%
\bibitem [{\citenamefont {Akhiezer}\ and\ \citenamefont
  {Kemmer}(1965)}]{Akhiezer1965}%
  \BibitemOpen
  \bibfield  {author} {\bibinfo {author} {\bibfnamefont {N.I.}\ \bibnamefont
  {Akhiezer}}\ and\ \bibinfo {author} {\bibfnamefont {N.}~\bibnamefont
  {Kemmer}},\ }\href@noop {} {\emph {\bibinfo {title} {{The classical moment
  problem: and some related questions in analysis (Vol. 5)}}}}\ (\bibinfo
  {publisher} {Oliver {\&} Boyd},\ \bibinfo {address} {Edinburgh},\ \bibinfo
  {year} {1965})\BibitemShut {NoStop}%
\bibitem [{\citenamefont {Alonso}\ and\ \citenamefont
  {Gorin}(2016)}]{Alonso2016}%
  \BibitemOpen
  \bibfield  {author} {\bibinfo {author} {\bibfnamefont {L}~\bibnamefont
  {Alonso}}\ and\ \bibinfo {author} {\bibfnamefont {T}~\bibnamefont {Gorin}},\
  }\bibfield  {title} {\enquote {\bibinfo {title} {{Joint probability
  distributions for projection probabilities of random orthonormal states}},}\
  }\href {\doibase 10.1088/1751-8113/49/14/145004} {\bibfield  {journal}
  {\bibinfo  {journal} {Journal of Physics A: Mathematical and Theoretical}\
  }\textbf {\bibinfo {volume} {49}},\ \bibinfo {pages} {145004} (\bibinfo
  {year} {2016})},\ \Eprint {http://arxiv.org/abs/arXiv:1510.05333v1}
  {arXiv:arXiv:1510.05333v1} \BibitemShut {NoStop}%
\end{thebibliography}%

\end{document}